# Direction-Dependent Lateral Domain Walls in Ferroelectric Hafnium Zirconium Oxide and their Gradient Energy Coefficients: A First Principles Study

Tanmoy Kumar Paul*, Atanu Kumar Saha, Sumeet Kumar Gupta

*Purdue University, West Lafayette, Indiana, 47907, USA*

Email: paul115@purdue.edu / Phone: (765) 607-3147

To understand and harness the physical mechanisms of ferroelectric Hafnium Zirconium Oxide (HZO)-based devices, there is a need for clear understanding of domain interactions, their dynamics, negative capacitance effects, and other multi-domain characteristics. These crucial attributes depend on the coupling between neighboring domains quantified by the gradient energy coefficient (*g*). Furthermore, HZO has unique orientation-dependent lateral multidomain configurations. To develop an in-depth understanding of multi-domain effects, there is a need for thorough analysis of *g*. In this work, the energetics of multidomain configurations and domain growth mechanism corresponding to lateral domain walls of HZO are analyzed and gradient energy coefficients are quantified using first-principles Density Functional Theory calculations. These results indicate that one lateral direction exhibits the following characteristics: i) DW is ultra-sharp and domain growth occurs unit-cell-by-unit-cell, ii) the value of *g* is negative and in the order of $10^{-12}\,\text{Vm}^3\text{C}^{-1}$, and iii) *g* reduces (increases) with compressive (tensile) strain. In contrast, in the other lateral direction, the following attributes are observed: i) DW is gradual and domain growth occurs in quanta of *half*-unit-cell, ii) *g* is positive and in the order of $10^{-10}\,\text{Vm}^3\text{C}^{-1}$, and iii) *g* increases (reduces) with compressive (tensile) strain.

## 1. Introduction

Ferroelectric (FE)-based devices such as FE Tunnel Junction (FTJ), FE Field Effect Transistor (FEFET), and FE heterojunctions have shown immense promise for various applications, including non-volatile memory, logic, and neural networks, with new applications being continually explored. [1] [2] [3] With ongoing efforts of device miniaturization, HfO2-based materials (both doped and undoped) have significantly intrigued the FE research community. [4] Among HfO2 based FEs, 50% Zr-doped HfO2, widely known as Hafnium Zirconium Oxide (HZO) shows enhanced remnant polarization (experimental value of 23 μC cm$^{-2}$) and stable ferroelectricity in ultra-thin samples. [5] [6] In addition, HZO exhibits strong multi-domain (MD) effects.[7] For understanding the physics of domain interaction in MD FEs, the phase-field modeling approach has been an instrumental tool. [8] It utilizes the Ginzburg-Landau-Devonshire (LGD) equation to describe the FE state of the material via polarization. [9] According to this theory, the domain configuration in FE is the result of a complex interplay among several energy components, i.e., free energy, depolarization energy, gradient/elastic energy, and others, as described by Equation (1).

$$f_{\text{total}} = f_{\text{free}} + f_{\text{grad}} + f_{\text{dep}}$$

$$f_{\text{free}} = f_0 + \frac{\alpha}{2}P_z^2 + \frac{\beta}{4}P_z^4 + \frac{\gamma}{6}P_z^6$$

$$f_{\text{dep}} = -E_z P_z$$

$$f_{\text{grad}} = \frac{g_x}{2}\left(\frac{\partial P_z}{\partial x}\right)^2 + \frac{g_y}{2}\left(\frac{\partial P_z}{\partial y}\right)^2 + \frac{g_z}{2}\left(\frac{\partial P_z}{\partial z}\right)^2 \quad (1)$$

Here, $P_z$ is the polarization along the *z*-axis (assumed to coincide with the polar *c*-axis of the *Pca2₁* unit cell without any loss of generality), $f_{\text{free}}$ is the local free energy density, $f_{\text{grad}}$ is the gradient energy density, and $f_{\text{dep}}$ is the depolarization energy density which results from bound polarization charges near surface/interface, α, β and γ are Landau's free energy coefficients, $E_z$ is the electric field in FE along *z*, and $g_x$, $g_y$ and $g_z$ are gradient energy coefficients along *x*, *y*, and *z* axes, respectively. Gradient energy is related to the formation of domain wall (DW) and is defined as the associated energy cost owing to the polarization gradient near the DW region. Mathematically, $g_x$, $g_y$, and $g_z$ are proportionality constants that determine the magnitude of gradient energy contribution to the total energy.

In the absence of bound charge compensation at the interface, depolarization fields emerge inside the FE leading to an increase in $f_{\text{dep}}$ (see Equation 1) and thus, the overall energy. [10] To reduce $f_{\text{dep}}$, the FE material may break into multiple domains with lateral DWs. In this case, interface bound charges are partially compensated via an in-plane stray field between two neighboring domains. With a higher density of alternating domains, increased localization of stray field and more suppression of depolarization field occurs. However, this comes at the cost of net increased gradient energy governed by the coefficient, *g*. Ultimately, the two opposing factors are balanced by each other to minimize the overall energy. This leads to the stabilization of the most energetically favored

domain configuration according to the LGD formalism.[11]

Earlier works based on phase field models have demonstrated the importance of $g$ for a wide number of device phenomena in Hafnium Zirconium Oxide (HZO) and is summarized in Figure 1.[12] For example, gradient energy dictates the domain density (which increases with decrease in $g$). This, in turn, determines whether the polarization switching is dominated by domain nucleation (typically for low domain density) or DW displacement (associated with high domain density). Besides, $g$ also impacts the softness/hardness of the DW (sharpness of polarization variation). Small $g$ results in hard DW whereas large $g$ results in soft DW, which are associated with hysteretic and non-hysteretic characteristics, respectively. Thus, $g$ determines the suitability of the FE device for memory or logic applications.[13] Furthermore, the negative capacitance effect in FEs, which has shown intriguing characteristics and promising applications, is also closely tied to multi-domain behavior and hence, to the value of $g$.[14][15] It is, thus, clear that a better understanding of $g$ is highly significant to advance the exploration and design of FE-based devices and circuits.

Being crucial for FE device phenomena, the efforts to quantify $g$ in traditional perovskite-based FEs are many. For example, by combining first-principle calculations and hyperbolic function fitting, $g_x$, $g_y$, and $g_z$ in PbTiO3 are reported as $0.12\times10^{-10}$, $0.12\times10^{-10}$, and $0.21\times10^{-10}$ Vm$^3$C$^{-1}$ respectively.[16] Using density functional perturbation theory and long wavelength expansion of crystal Hamiltonian, $g_x$ of six perovskite oxides (BaTiO3, CaTiO3, KNbO3, NaNbO3, PbTiO3, and PbZrO3) have been calculated which fall in the range of $0.2\times10^{-10}$ - $1\times10^{-10}$ Vm$^3$C$^{-1}$.[17] On the other hand, for FE HfO2, negligible $g$ has been predicted due to flat phonon bands of polar transverse optic phonons.[18] However, proper quantification of $g$ in different spatial directions is still lacking in HfO2-based FEs.

Being hitherto less explored, it is crucial to quantify $g$ associated with 180° lateral DWs in HZO. This is because FE MD configurations in HZO are primarily 180° lateral domains where the DW plane is parallel to the polarization axis of the unit cell.[19] Further, the unit cell of the FE phase in HZO has a unique type of atomic configuration. In one of the lateral directions, unit cells can be divided into two layers: a non-polar dielectric (DE) layer and a FE layer.[20] In the other lateral direction, continuous FE layers are formed. Due to this, domain width, polarization, crystal phase at the DWs and their width, etc. are highly dependent on the crystal orientation of the domains. Correspondingly, the gradient energy associated with the former (alternate

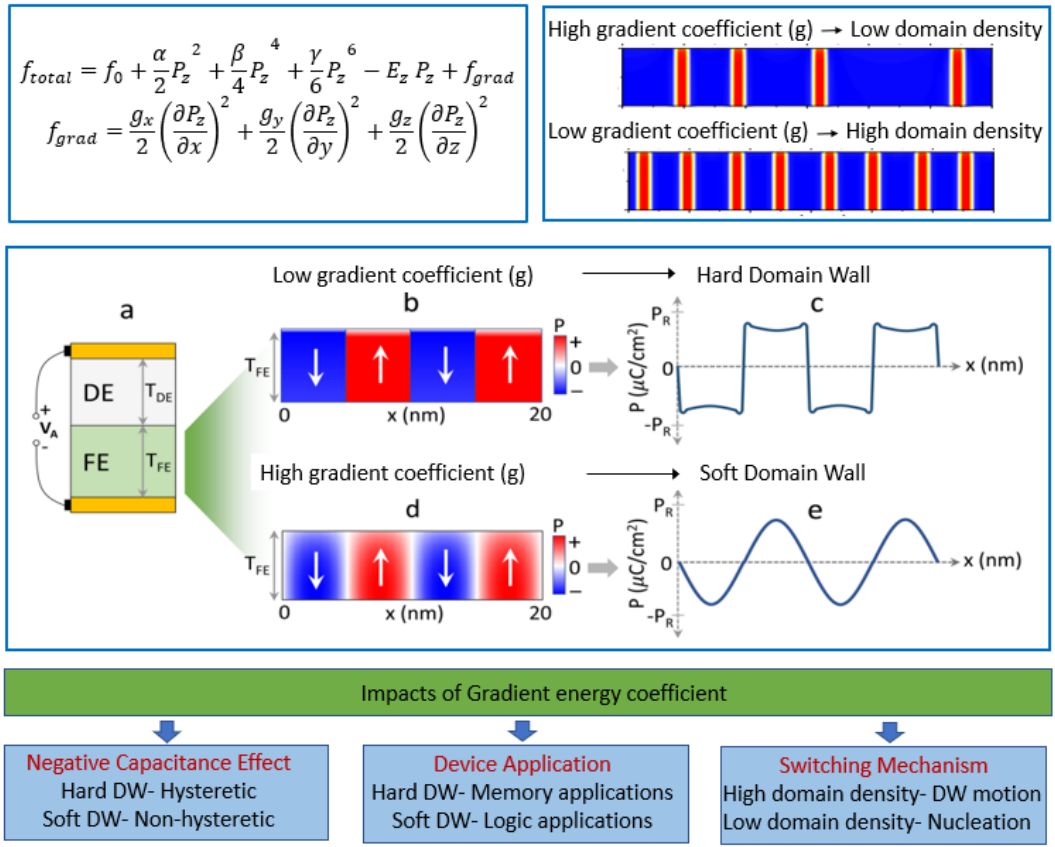

**Figure 1** Mathematical representation of gradient energy coefficient and its significance. The chart shows how the gradient energy coefficient impacts several ferroelectric device properties and leads to device-level applications.

FE-DE) DW is expected to be quite different from the latter (continuous FE) DW.

In this work, we comprehensively quantify the $g$ parameter in FE HZO considering its direction-dependent material physics using first-principles Density Functional Theory (DFT) calculations. Specific contributions are as follows:

➢ We find atomic-scale strain-free (lattice matched) bulk MD configurations in both the lateral directions (i.e., $x$ and $y$ directions which are perpendicular to the polarization direction).

➢ We calculate $g_x$ and $g_y$ for the most energetically stable lateral DWs assuming constant $g$ along a particular direction.

➢ We present a detailed quantitative comparison of the direction dependence of $g$.

➢ We explore the microscopic switching mechanism in both lateral directions to understand their domain kinetics.

➢ We analyze the strain dependence of $g$ to predict its characteristics in the cases when HZO is integrated into a device and is subjected to strain due to electrodes/substrates.

## 2. Results and Discussion

To obtain the MD configurations of HZO, we perform extensive first-principles calculations based on DFT. The details of the computational procedure are described in the Computational Section. Here, let us first introduce the unit cell of HZO. Both experimental and theoretical works have proved that the orthorhombic $Pca2_1$ phase is the source of ferroelectricity in HfO$_2$-based FE materials. [21] [22] The unit cell of FE Hf$_{0.5}$Zr$_{0.5}$O$_2$ is like the $Pca2_1$ phase of HfO$_2$ where two Zr dopant atoms replace Hf at any two of the four lattice sites. We consider various cases of the lattice site occupation by Hf/Zr atoms. Different permutations of Hf and Zr can be narrowed down to three primitive configurations. We depict these three cases in **Figure 2**a-c where each plane parallel to (100), (010), or (001), respectively, contain, a single type of atom; either Hf, Zr, or O atoms. Alternately, we can say that planes perpendicular to one of the three orthogonal directions are mono-atomic. Among the three configurations, the formation of mono-atomic planes perpendicular to the [001] direction (i.e., along the polarization or $z$ direction) in Figure 2c yields the minimum energy. Note that, the maximum deviation of the relaxed energy among the three cases (≈4 meV per unit cell) is very small. Nevertheless, unless mentioned otherwise, we subsequently consider this minimum energy configuration as the HZO unit cell.

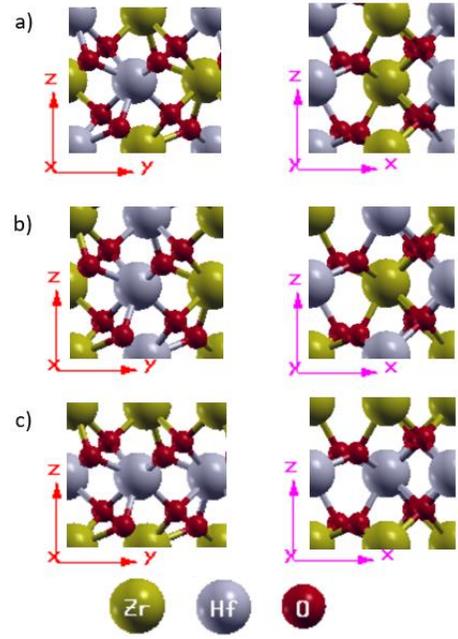

**Figure 2** Different lattice sites occupation by Hf/Zr atoms in Orthorhombic Hf$_{0.5}$Zr$_{0.5}$O$_2$ unit cells. Planes parallel to a) (100), b) (010), and c) (001) are monoatomic, respectively. Among them, (c) gives the minimum energy configuration. Note that [001] is the polarization direction. Golden, Gray, and red atoms are Zr, Hf, and O atoms, respectively. For all other figures, both Zr and Hf atoms are colored in gray.

The calculated relaxed lattice parameters of the HZO unit cell (5.05, 5.28, and 5.08 A° along the three orthogonal directions) match with previous theoretical calculations and experiments. [23] [5] We obtain spontaneous polarization of 50 μC cm$^{-2}$ with polarization quantum of 120.24 μC cm$^{-2}$ along the $z$-direction. The spontaneous polarization obtained here is higher than experimentally observed residual polarization. [5] This is attributed to the polycrystalline nature of FE HZO, which leads to lower average polarization in the experiments. But, the polarization from our simulations matches well with previous theoretical calculations. [20]

For gradient energy calculations, ionic relaxations of MD supercells are performed using the lattice parameters of the relaxed single-domain (SD) cell. We define these SD and MD supercells as unconstrained SD and constrained MD supercells, respectively. Later, during analyzing results, we also perform calculations on the fully relaxed MD supercells and define them as unconstrained MD supercells. MD supercells of different sizes represent domains of different widths.

Once the SD and MD configurations are optimized, we obtain the DW energy of MD configurations. DW energy is widely used to compare the favorability of MD configurations over SD, as well as among several MD configurations. It is defined as the energy difference between MD ($E_{MD}$) and SD ($E_{SD}$) supercell per unit cross-sectional area of each DW, as shown in Equation (2).

$$\text{DW energy, } E_{DW} = \frac{E_{MD} - E_{SD}}{2A} \quad (2)$$

Here, factor 2 in the denominator indicates that there are two DWs in each periodic supercell and $A$ is the cross-sectional area of each DW.

For a particular type of lateral DW, we obtain the configuration with the lowest DW energy and calculate the gradient energy coefficient for that configuration. As described by Equation (1), gradient energy density depends on the polarization profile. Bulk structures are infinitely periodic, and no surface/interface or polarization-bound charges exist. Hence, depolarization energy does not take part in the total energy. Thus, in the SD supercell, total energy ($E_{SD}$) contains only the free energy component. On the contrary, for the MD supercell, both local free energy and gradient energy contribute to the total energy ($E_{MD}$). The energy barrier profile from the NEB method along with the modern theory of polarization provides the functional dependence of local free energy on polarization. Thus, the calculated polarization profile yields the local free energy of each dipole in MD ($E_{local,MD}$) and SD ($E_{local,SD}$) supercells. We use these values in Equation (3) to calculate the gradient energy density ($f_{grad}$).

Total energy difference per unit volume,

$$\frac{E_{MD} - E_{SD}}{V} = \frac{1}{V}\Sigma(E_{local,MD} - E_{local,SD}) + f_{grad} \quad (3)$$

Then, the local polarization profile quantifies the corresponding gradient energy coefficient ($g$) as described in Equation (4).

$$f_{grad} = \Sigma \frac{1}{2} g \left(\frac{\Delta p}{\Delta l}\right)^2 \quad (4)$$

Here, $\Delta l$ is the distance between two consecutive dipoles and $\Delta p$ is the difference in polarization between the dipoles. The method followed here to calculate the gradient energy coefficient is like the method followed previously for perovskites, except the fact that we extract all values from first principles calculations. [16] Based on the discussion in Section 1, we know that DWs in HZO have significant directional properties. Now, let us delve deeper into the emergence of different DWs in different directions and their energetics.

## 2.1 Domain Wall Configurations

The orthorhombic unit cells of HZO shown in **Figure 3** are composed of two layers along [100] direction: a DE/spacer layer with all the O atoms in the

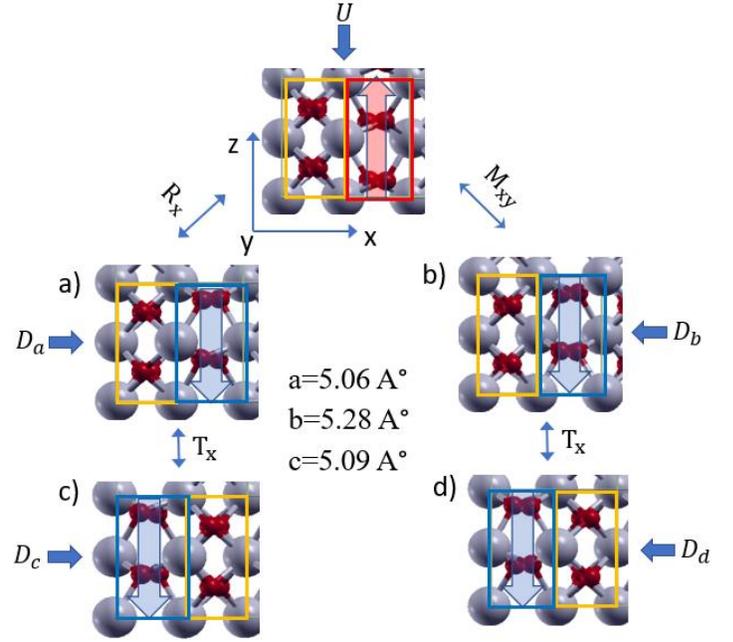

**Figure 3** Orthorhombic $Hf_{0.5}Zr_{0.5}O_2$ unit cells. For an upward polarized unit cell in the top, four energetically equivalent downward polarized unit cells in a-d) are derived from Euclidean transformations ($R_x$ = rotation with respect to x-axis, $M_{xy}$ = reflection in xy plane, and $T_x$ = Translation along x-axis). We will denote the upward polarized unit cell as $U$ and the four downward polarized unit cells as $D_a$, $D_b$, $D_c$, and $D_d$, respectively. Polarization is along the $z$ direction. Blue and red outlined regions are non-centrosymmetric polar layers with downward and upward polarizations, respectively. Yellow outlined regions are centrosymmetric spacer layers. The relaxed lattice parameters of all the unit cells are identical and denoted in the figure.

centrosymmetric position and FE/polar layer with all the O atoms in the non-centrosymmetric position. We will call this "Alternate Polar Spacer Layer" (APSL) configuration. The spacer and polar layers are indicated by the yellow and red/blue outlined regions, respectively in Figure 3. This type of spacer layer in between the polar layers is unique to HZO, and fundamentally different from Perovskite-based conventional FEs. [18] Depending on the relative position of these spacer and polar layers, as well as their atomic co-ordination, different types of atomic arrangements are possible in an HZO unit cell. [24] For example, for a given upward polarized unit cell, four types of downward polarized unit cells with different atomic arrangements are derived by performing Euclidean transformations i.e. rotation with respect to the $x$ axis ($R_x$), reflection in $xy$ plane ($M_{xy}$) and translation along the $x$ axis ($T_x$) (Figure 3a-d). We will denote the upward-polarized unit cell as $U$ and the 4 downward-polarized unit cells as $D_a$, $D_b$, $D_c$, and $D_d$ respectively. All the unit cells are energetically identical

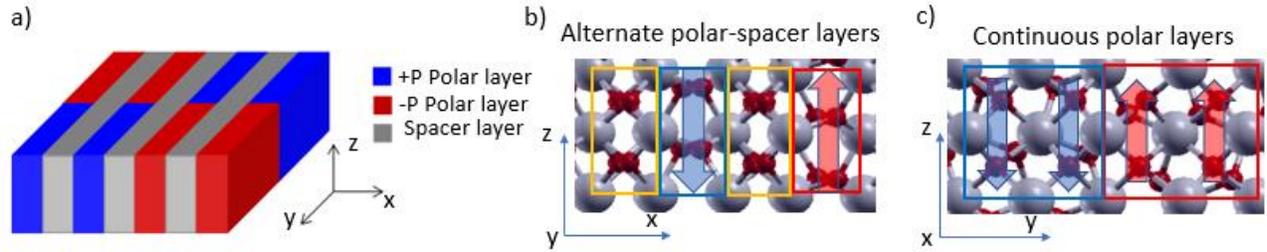

**Figure 4** a) Simplistic 3D view of lateral domain walls in orthorhombic HZO. Alternate polar and spacer layers are along the *x* direction for both domains and along the *y* direction all layers are originally polar in nature. Atomic configurations of b) alternate polar-spacer layers (APSL) and c) continuous polar layers (CPL). Blue and red outlined regions are non-centrosymmetric polar layers with downward and upward polarization, respectively. Yellow outlined regions are centrosymmetric spacer layers. (DWs in the figure are before relaxation and for understanding purposes only.)

in the SD configuration and their atomic coordinates are explained in Supplementary Figure S1. There is an inherent energy barrier involved in flipping the polarization of a unit cell. The energy barrier pathways from upward-polarized unit cell to each of the downward-polarized unit cells are included in Supplementary Figure S2.

Unlike [100], along [010] direction each layer is polarized as can be seen from Supplementary Figure S3(b). This is because each layer along [010] contains both non-centrosymmetric and centrosymmetric oxygen atoms. We will call this "Continuous Polar Layer" (CPL) configuration. Thus, domains with APSL or CPL configurations in HZO are different from each other in FE nature. Hence, depending on the direction of the MD formation, DWs in HZO have different characteristics as well. As a result, corresponding gradient energies are fundamentally distinct in magnitude. A simplistic 3D view of an HZO grain with the two types of lateral DW is shown in **Figure 4**a to depict the directional dependence. Figure 4b and 4c illustrate the corresponding atomic configurations along each lateral direction before MD relaxation. Although in real FE grains, there might be both types of DWs simultaneously, here, we consider them separately so that the energy state of each type of DW can be obtained individually. Thus, our analysis does not consider the energy contribution due to the interaction between the APSL and CPL DWs.

Now, to find the minimum energy MD configuration for each type of DWs, we calculate all the combinations of upward and downward-polarized unit cells shown in

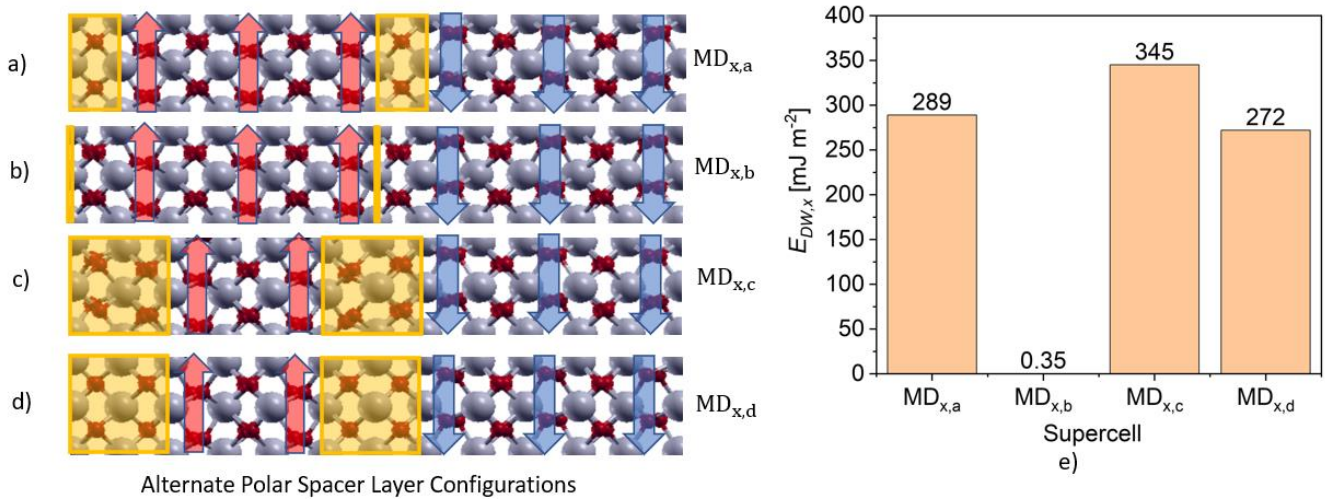

**Figure 5** Representative 6-unit supercell for four possible APSL DWs along [100] direction, a) $MD_{x,a}$, b) $MD_{x,b}$, c) $MD_{x,c}$, and d) $MD_{x,d}$ deriving from combinations of the upward polarized unit cell, *U* with downward polarized unit cells, $D_a$, $D_b$, $D_c$, and $D_d$, respectively. The atomic configurations in DW resembles tetragonal, orthorhombic, distorted orthorhombic, and tetragonal, respectively. e) Domain Wall energy comparison of each type of APSL DW. For proper comparison purposes, all supercells comprise an equal number of unit cells. Yellow-shaded regions are DW regions for each case. All MD supercells are obtained from ionic relaxation at equivalent SD lattice parameters for representing SD to MD transition. $MD_{x,b}$ is the most energetically favorable APSL DW and is used for future calculations on APSL.

Figure 3. For APSL DW, $U$ combines with $D_a$, $D_b$, $D_c$, and $D_d$ to form MD configurations $MD_{x,a}$, $MD_{x,b}$, $MD_{x,c}$, and $MD_{x,d}$, respectively ($x$ denotes APSL direction) as shown in **Figures 5**a-d. The corresponding DW energies are compared in Figure 5e. DW energy is very small (< 1 mJ m$^{-2}$) in $MD_{x,b}$ where the spacer layer acts as an in-built wall between the domains. This is because the crystal phase near the DW is nearly orthorhombic *Pca2₁* and almost identical to the low-energy bulk atomic configuration deep inside the domains. $MD_{x,a}$ and $MD_{x,d}$ show the tetragonal phase in the spacer layers near the DW. As the orthorhombic atomic arrangement is lost near the DWs, the structural change causes DW energies to be higher than that of $MD_{x,b}$. Among the two, $MD_{x,d}$ has been denoted as topological DW for low barrier switching in earlier works. [24] $MD_{x,c}$ shows atomic distortions near the DW suggesting the highest DW energy amongst the four configurations. Thus, $MD_{x,b}$ is the most energetically favorable DW and has been considered for subsequent analysis. The atomic configurations near all the DWs are included in Supplementary Figure S4.

Similarly, for CPL DW, combinations of $U$ with $D_a$, $D_b$, and $D_c$ generate MD configurations $MD_{y,a}$, $MD_{y,b}$, and $MD_{y,c}$ respectively ($y$ denotes CPL direction) as shown in **Figure 6**a-c. A combination of $U$ and $D_d$ gives unstable MD and so, we have discarded it for energy calculations. $MD_{y,a}$ and $MD_{y,c}$ show distorted tetragonal-like phases at the center of the DW. In $MD_{y,b}$, the atomic configurations in the DW show *pbcm*-like phase.

The atomic configurations near all the DWs are included in Supplementary Figure S5. DW energies denoted in Figure 6d show that $MD_{y,a}$ is the most energetically stable DW and will be considered for further calculations of CPL DW. It is noteworthy that the atomic configuration near CPL DW is significantly different from the low-energy bulk atomic configuration deep inside the domains. So, in the most energetically favorable configurations, CPL DW energy is much larger compared to the APSL DW.

## 2.2 Gradient Energy Coefficient

### 2.2.1 APSL Domain Wall

A representative relaxed MD supercell in the APSL configuration consisting of domains of equal widths (6 upward and 6 downward polarized unit cells) is shown in **Figure 7**a. The APSL DW is atomically sharp as can be seen from the unit cell based local polarization profile in Figure 7b and can be treated as a hard DW. As stated earlier, the in-built spacer layer in between the two domains makes it possible to orient the two neighboring polar layers in opposite directions, thus creating an abrupt DW. This is consistent with previous research work. [18] The polarization is uniform in each domain except for a mild increase near the DW (by ≈4 μC cm$^{-2}$ in magnitude compared to the polarization inside the domain). We have observed abruptness and mild enhancement of polarization near the DW irrespective of the domain width (i.e., supercell size). The mild polarization increase originates as follows. When polarization reversal of a domain occurs through the *pbcm* phase, the non-centrosymmetric oxygen atoms move away from the centrosymmetric position toward the Hf/Zr plane (Supplementary Figure S2b). In section 2.3.1 we will show that this is the polarization reversal path MD prefers energetically. We find that in the relaxed static APSL configuration, the atomic arrangement in the unit cells near the DW is slightly deviated from the *Pca2₁* phase toward the *pbcm* phase. Thus, the polarization (calculated with respect to the centrosymmetric $Fm\bar{3}m$ phase) is slightly enhanced compared to *Pca2₁* phase (Supplementary Figure S6a).

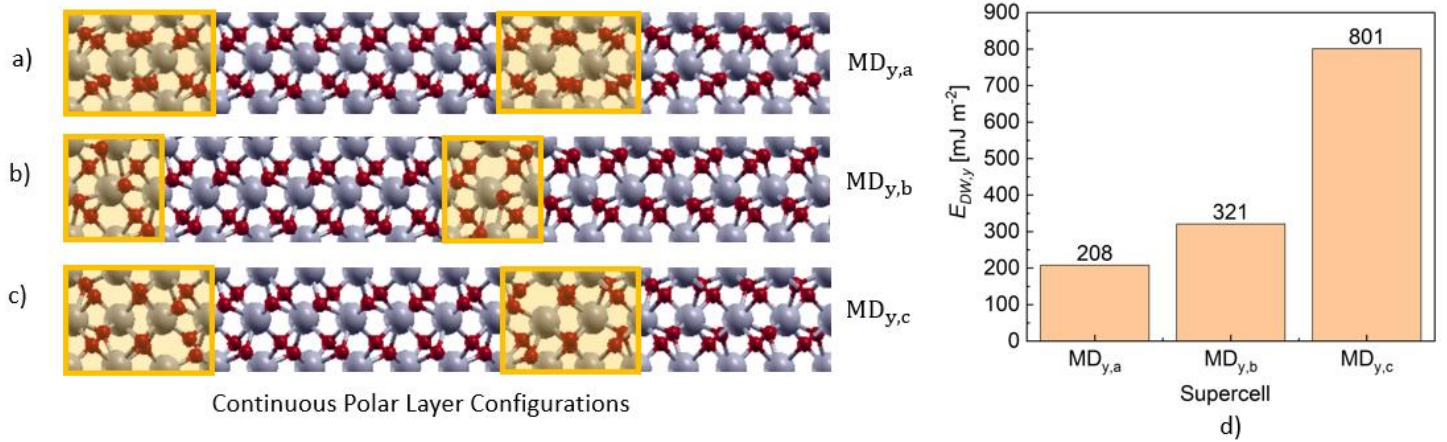

**Figure 6** Representative 8-unit supercell for three possible CPL DWs along [010] direction, a) $MD_{y,a}$, b) $MD_{y,b}$, and c) $MD_{y,c}$ deriving from combinations of the upward polarized unit cell, $U$ with downward polarized unit cells $D_a$, $D_b$, and $D_c$, respectively. (Combination with $D_d$ gives unstable configuration.) The atomic configurations at the center of the DW resemble distorted tetragonal, *pbcm*, and distorted tetragonal, respectively d) Domain Wall energy comparison of each type of CPL DW. For proper comparison purposes, all supercells comprise an equal number of unit cells. Yellow-shaded regions are DW regions for each case. All MD supercells are obtained from ionic relaxation at equivalent SD lattice parameters for representing SD to MD transition. $MD_{y,a}$ is the most energetically favorable CPL DW and is used for future calculations on CPL.

Now let us look at the dependence of DW energy on domain width shown in Figure 7c. In APSL, unconstrained MD shows slight reduction in supercell volume compared to SD as seen in other theoretical studies of $HfO_2$.[18] So, during constrained transition from SD to MD with fixed lattice parameter, the MD supercell feels compressive strain. As domain width increases, the magnitude of compressive strain reduces as illustrated in Supplementary Figure S7(a). Hence, with wider domains, the constrained MD supercells are closer to the corresponding minimum energy unconstrained MD volume. As a result, DW energy decreases with increasing domain width and becomes negative for wider domains. However, the magnitude of DW energy is always very small (< 1 mJ m⁻²).

With this DW energy, the calculated gradient energy coefficient in APSL ($g_x$) is shown in Figure 7d. Here, the polarization gradients between consecutive unit cells are considered for the calculation of $g_x$. This is because, in HZO, the existence of spacer and polar layers is intrinsically coupled. So, polarization variation within unit cell does not affect the gradient energy in APSL. The local free energies of each of the unit cells in the supercell are extracted from the energy barrier profile in Supplementary Figure S6(a). The calculated result indicates three important attributes: i) $g_x$ is negative, ii) its absolute magnitude is very small (order of $10^{-12}$ V m³ C⁻¹), and iii) absolute magnitude of $g_x$ decreases with increasing domain width.

These attributes can be explained as follows. The local free energy of the unit cells within the domains of MD are equal to those of SD because of their identical polarizations and atomic configurations. Moreover, since the DW energy is negligible, Equation (3) suggests that gradient energy is almost fully determined

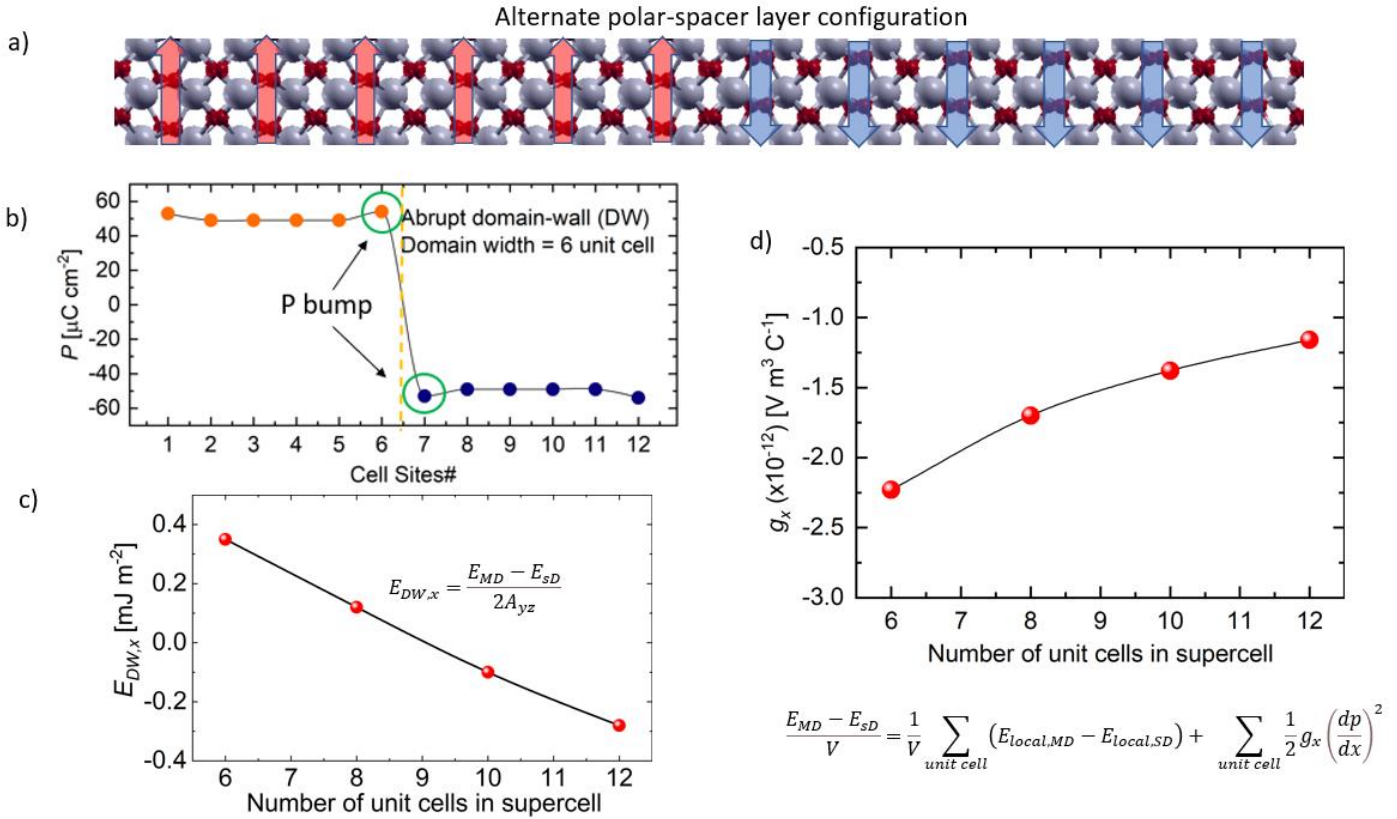

**Figure. 7** a) Relaxed symmetric 12-unit supercell of APSL DW. b) Unit cell-based local polarization profile showing an atomically sharp DW. Polarization bump is observed near the DW. c) DW energy variation and d) Gradient energy coefficient with domain size. Figures (c) and (d) contain the equations used for calculations. Unit cell polarization is used because of the coupled existence of polar layers and spacer layers throughout the domains and DWs. Local free energies are extracted from the energy barrier profile for polarization reversal of unit cells. The DW energy is negligible which signifies the orthorhombic-like nature of the DW. The gradient energy coefficient is negative and is mainly determined by the local free energy density.

by local free energy contribution of the unit cells surrounding the DWs. Near the DW, the unit cells of MD supercell are slightly deviated from the minimum energy point of energy barrier profile as described in the first paragraph of this section. So, $E_{\text{local,MD}}$ is always slightly $> E_{\text{local,SD}}$. Thus, $g_x$ becomes negative according to the equation shown in Figure 7d. It turns out that the absolute magnitude of gradient energy coefficient ($g_x$) is small and of the order of $10^{-12}$ V m$^3$ C$^{-1}$ (two orders less than that of conventional perovskite based FEs). [17]

To show the trend of $g_x$ with domain width, the local energy of unit cells near DW for different sized supercells are extracted from the barrier profile. As stated earlier, the polarization profile is similar irrespective of supercell size. Thus, we obtain similar local energy distribution for different supercell sizes. The net local energy difference being the same, an increase in the domain width reduces the difference in local energy density. Hence, $|g_x|$ reduces with domain width as shown in Figure 7d.

To use the extracted parameters in phase field models of HZO, one needs to account for the polarization bump at the two lattice points near the DW in addition to the negative value of $g_x$. Previous theoretical analysis has shown that higher order polarization gradient ($f_{\text{grad,high}} = \frac{1}{2} w_x \left(\frac{d^2 P}{dx^2}\right)^2$) can have significant contribution to the total energy in addition to the contribution of $g$ in ultra-sharp 180° DWs. [25] Here, $w_x$ is the higher order gradient energy coefficient along APSL direction. Mathematically, the higher order gradient energy for ultra-sharp DW has dominant effect to the energy component at the two lattice points near the DW. Thus, we equate $f_{\text{grad,high}}$ to the excess local free energy at those two lattice points and estimate the value of $w_x$. The calculation method and calculated $w_x$ parameter is discussed in Supplementary Figure S8.

### 2.2.2 CPL Domain Wall

A representative relaxed MD supercell with equally wide domains for CPL is shown in **Figure 8**a. Here, DW

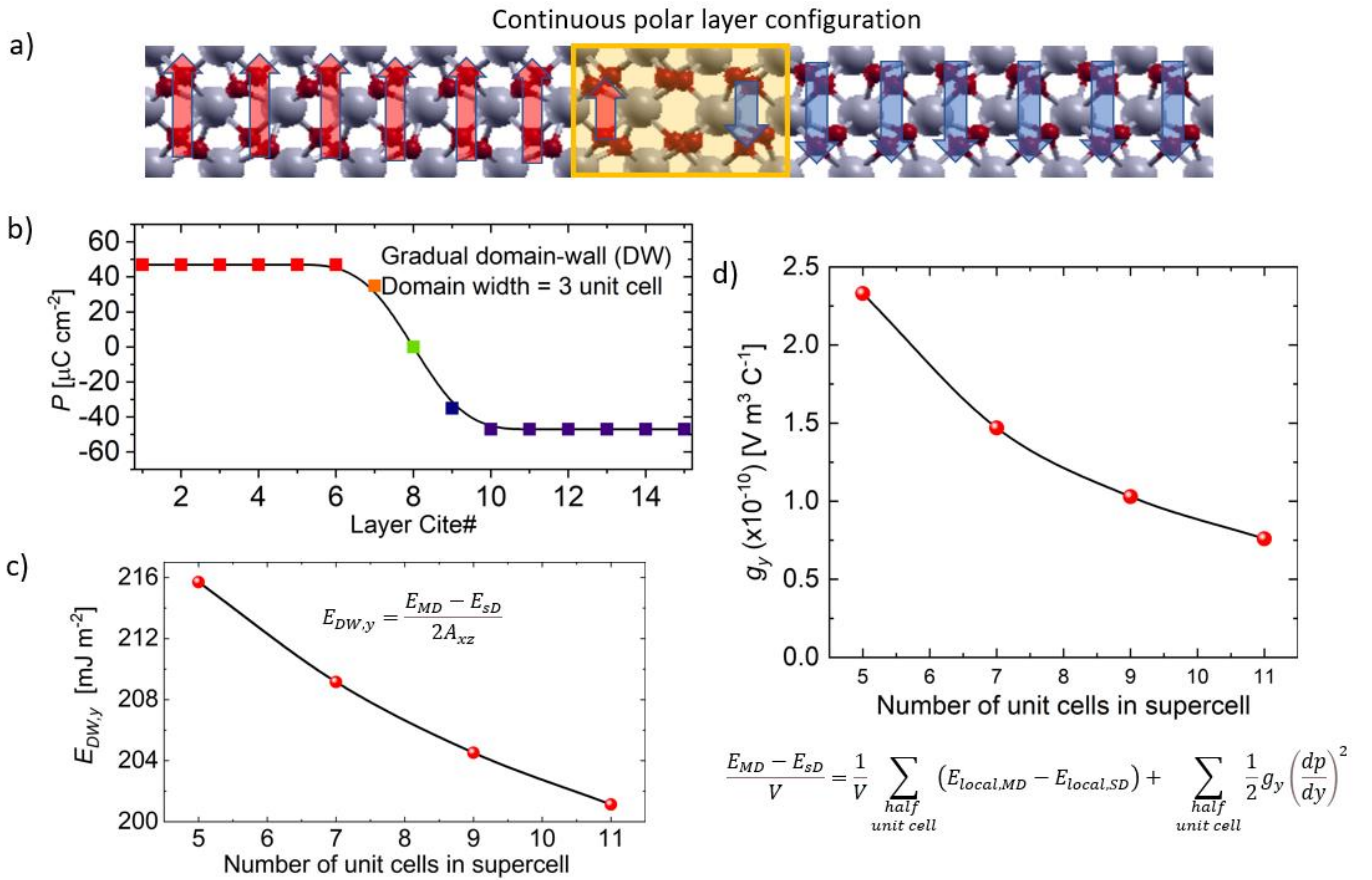

**Figure 8** a) Relaxed symmetric supercell of CPL DW. b) Half-unit-cell-wise local polarization profile showing a gradual DW. The distorted tetragonal layer at the center of the DW has zero polarization and converges to bulk polarization toward the domains. c) DW energy and Gradient energy coefficient with domain size. Figures (c) and (d) contain the equations used for calculations. Half-unit-cell-wise polarization is used because domains can consist of an odd number of half-unit cells and DW consists of the yellow outlined one-and-a-half-unit cell. Local free energies are extracted from the energy barrier versus the polarization profile of corresponding unit cells. High DW energy signifies different atomic configurations near the DW compared to low energy orthorhombic configurations deep inside the domains.

consists of one-and-a-half-unit cell (yellow shaded region). Besides, domains can consist of odd number of half unit cells. Thus, contrary to APSL, we find it more useful to consider half-unit-cell wise polarization in CPL. The polarization profile in Figure 8b shows gradual change in polarization along the direction of CPL. The distorted tetragonal layer formed between the two domains has zero polarization and the polarization gradually reaches the bulk value as we move away from the DWs. This is different from the sharp polarization profile in the APSL configuration, but similar to that observed in perovskite based FEs. [18]

DW energy variation with domain width is shown in Figure 8c. In CPL, unconstrained MD shows slight increase in supercell volume compared to SD. So, during constrained transition from SD to MD with fixed lattice parameter, the MD supercell feels tensile strain. This tensile strain reduces with increase in the supercell size as described in Supplementary Figure S7(b). So, with increased width, the constrained MD supercells tend to move more toward the minimum energy unconstrained volume. As a result, DW energy decreases with increasing domain width.

For gradient energy coefficient calculation along CPL ($g_y$), half-unit-cell based polarization is used as shown in the equation in Figure 8d. The local free energy of each half-unit-cell in the supercell is extracted from the energy barrier profile in Supplementary Figure S6(b). Calculated $g_y$ is positive and in the order of $10^{-10}$ V m$^3$ C$^{-1}$. This is in the same order of what is obtained in perovskites. [17] The resultant $g_y$ reduces with increasing domain width following the trend of DW energy.

## 2.3 Nucleation and Domain Wall Motion

So far, we have analyzed the minimum energy static configurations of APSL and CPL DWs. Let us now look at the dynamics (nucleation of domains and DW motion) of both types of DW.

### 2.3.1 APSL Domain Wall

For the APSL configuration, nucleation of oppositely polarized domains and the growth of new domains correspond to the reversal of the polar layer. As the spacer layer always exists with the polar layer, each reversal step corresponds to one unit cell reversal. **Figure 9**a shows the domain growth mechanism. As discussed earlier, the DWs are atomically sharp, and either DW1 or DW2 shifts by one unit cell at each step. Due to the unit cell-by-unit cell polarization reversal, the energy barrier does not depend on the relative location of Hf/Zr atoms in the corresponding lattice sites.

Two paths of unit cell polarization reversal are possible from upward polarized unit cell, $U$ to downward polarized unit cell $D_b$ as discussed in Supplementary Figure S2. Therefore, for APSL MD configuration (consisting of $U$ and $D_b$ as discussed in Section 2.1), we consider two separate polarization reversal paths as shown in the top and bottom panels of Figure 9c. The path in the top panel corresponds to cubic $Fm\bar{3}m$ phase as the intermediate phase for polarization reversal. In this case, the non-centrosymmetric O atoms shift toward the other non-centrosymmetric position without crossing the Hf/Zr plane. The energy barrier for both nucleation and DW motion is almost equal in magnitude because of the abruptness of the DW. The barrier height is ≈1.34 eV which matches well with other theoretical calculations for HfO$_2$. [24] On the other hand, the path in the bottom panel corresponds to $pbcm$ as the intermediate phase for the reversal process. In this case, the non-centrosymmetric O atoms cross the Hf/Zr plane while reversing the polarization. Note that this path has a lower energy barrier of 0.76 eV (almost equal for nucleation and DW motion) and thus is energetically more favorable than the top one.

### 2.3.2 CPL Domain Wall

Now let us look at the domain growth mechanism in the CPL configuration. Here, stable domains with odd number of half-unit cells can exist. So, half-unit-cell-wise DW movement must be considered as shown in Figure 9b. The lengths of the arrow represent relative magnitude of the polarization. As discussed earlier for CPL, the DWs here represent gradually varying polarization. Here, the nucleation barrier is 640 meV, which is less than the APSL configuration, as well as the nucleation barrier of forming topological DW described for HfO$_2$. [24] After nucleation happens, any one of the two DWs can shift to grow the new domain. Here, due to the half-unit-cell-wise polarization reversal, the energy barrier depends on the relative location of Hf/Zr atoms in the corresponding lattice sites (corresponding to the three types of unit cell configurations in Figure 2a-c). Shifting of one DW might lead to a different barrier height than the other DW surrounding the nucleated domain. Supplementary Figure S9-S11 illustrates this dependence on the three cases showing different domain growth paths and barrier profiles. Here, we calculate the average of these energy barrier profiles for the growth of the two DWs. On average, the DW motion corresponds to 94 and 175 meV energy barriers for two consecutive layers. For the three lattice site occupation cases, average barrier heights are within 10 meV of the abovementioned values. Note that the DW growth barrier height in CPL is significantly less than that in APSL. According to

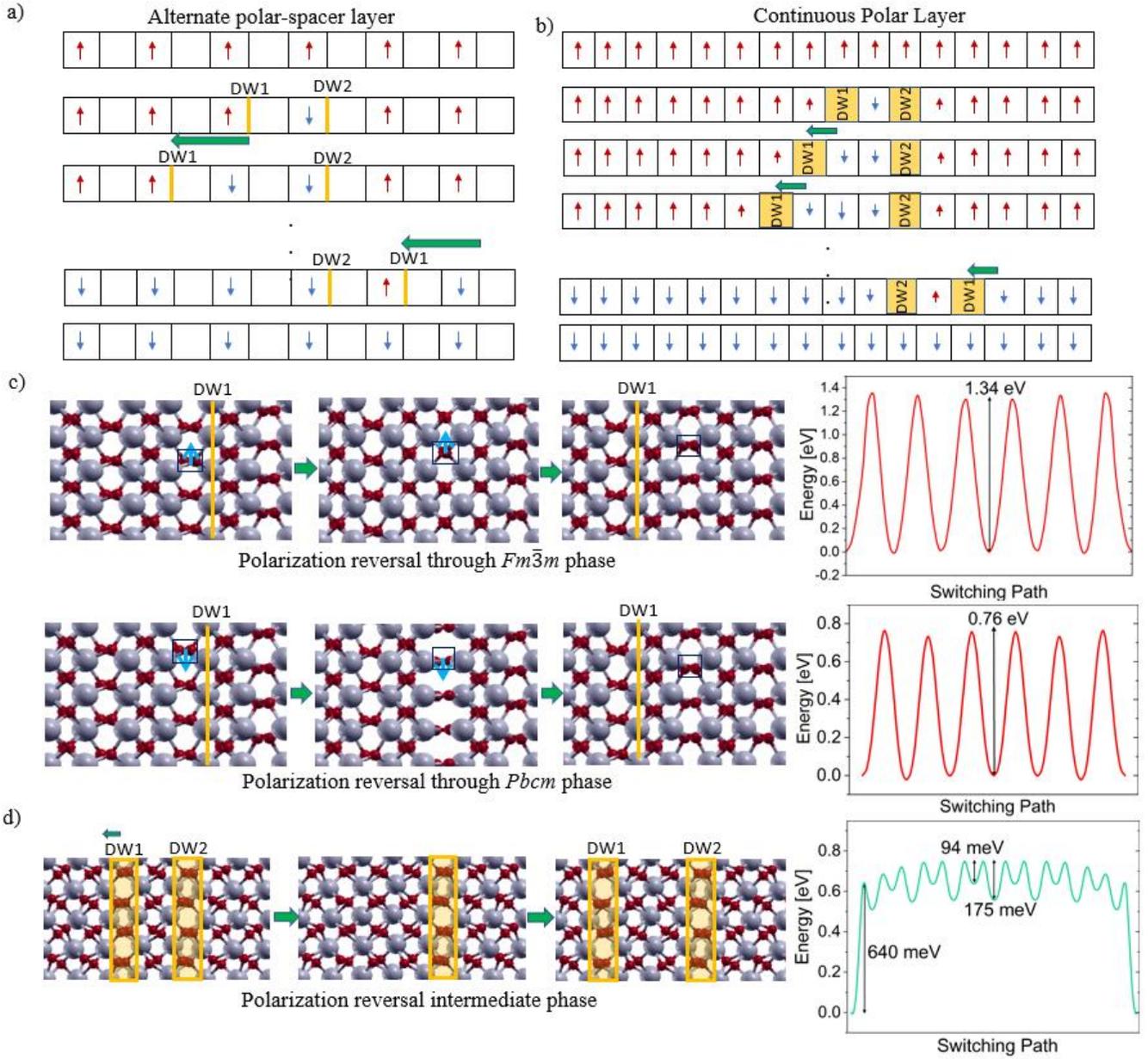

**Figure 9** Domain nucleation and DW motion mechanism for a) APSL and b) CPL configuration, respectively. Switching occurs by a unit cell and a half unit cell at each step, respectively. The length of the arrow defines the relative magnitude of polarization. c) Energy barrier profile for APSL domain growth mechanism. The barrier height is 1.34 eV when polarization reverses through the $Fm\bar{3}m$ phase and O atoms do not cross the Hf/Zr planes. On the contrary, a lower barrier height of 0.76 eV is obtained when O atoms cross the Hf/Zr plane during polarization reversal through the *pbcm* phase. Barrier heights are independent of the occupation site of Hf/Zr atoms, and the magnitude is similar for each reversal step. d) Average energy barrier profile of DW1 and DW2 shift for CPL domain growth mechanism. The barrier height for nucleation is 640 meV. DW motion barrier height for DW1 and DW2 is dependent on the lattice site occupation by Hf/Zr atoms. However, an average of 94 and 175 with 10 meV deviation is obtained in two consecutive steps for each of the three lattice site occupation cases.

Merz's law, DW speed is proportional to $\exp\left(-\frac{E_a}{E}\right)$, where $E_a$ is the activation field corresponding to barrier height and $E$ is the applied electric field. As a result, domain growth is expected to be much faster in CPL compared to APSL.

**2.4 Effect of Strain**

The stability of the orthorhombic phase in HZO is largely dependent on the strain. It has been shown in experiments that strain provided by electrodes and substrate helps stabilize the orthorhombic phases. [26] Therefore, it is important to understand its impact on $g_x$ and $g_y$. We will discuss this aspect in the current section.

### 2.4.1 APSL Domain Wall

As tensile strain is applied (**Figure 10**a), $g_x$ increases, i.e., it becomes less negative and approaches to positive value. On the other hand, for compressive strain, $g_x$ decreases, i.e., it becomes more negative.

To explain the trend in $g_x$ variation with strain, we calculate the DW energy for different strain conditions. Recall from section 2.2.1 that the APSL MD system feels slight compressive strain during the SD to MD transition. So, with applied compressive strain, the MD system energy initially reduces. However, after a certain critical compressive strain, it drifts away from the minimum energy. On the other hand, the SD system does not show such a non-monotonic behavior and exhibits an increase in energy with the increase in any compressive strain. So, DW energy decreases with compressive strain.

For applied tensile strain, constrained MD moves further away from its minimum energy monotonically. Although SD also moves away from its minimum energy, the deviation of MD is always larger than the deviation of SD from their respective minimum energy configurations at a particular tensile strain. As a result, with increasing tensile strain, DW energy increases. This dependence of DW energy on strain is shown in Figure 10b for a 6-unit supercell. As the magnitude of polarization is merely affected by strains up to 1%, gradient energy almost fully depends on the energy difference between SD and MD according to the Equation in Figure 7d. Thus, $g_x$ follows the same trend as the DW energy with strain.

### 2.4.2 CPL Domain Wall

In the case of CPL DW, tensile strain reduces $g_y$ while compressive strain enhances its value. This is shown in Figure 10c. Up to 1% tensile and compressive strains, $g_y$ remains in the order of $10^{-10}$ V m$^3$ C$^{-1}$. Now let us discuss the cause behind this trend in detail.

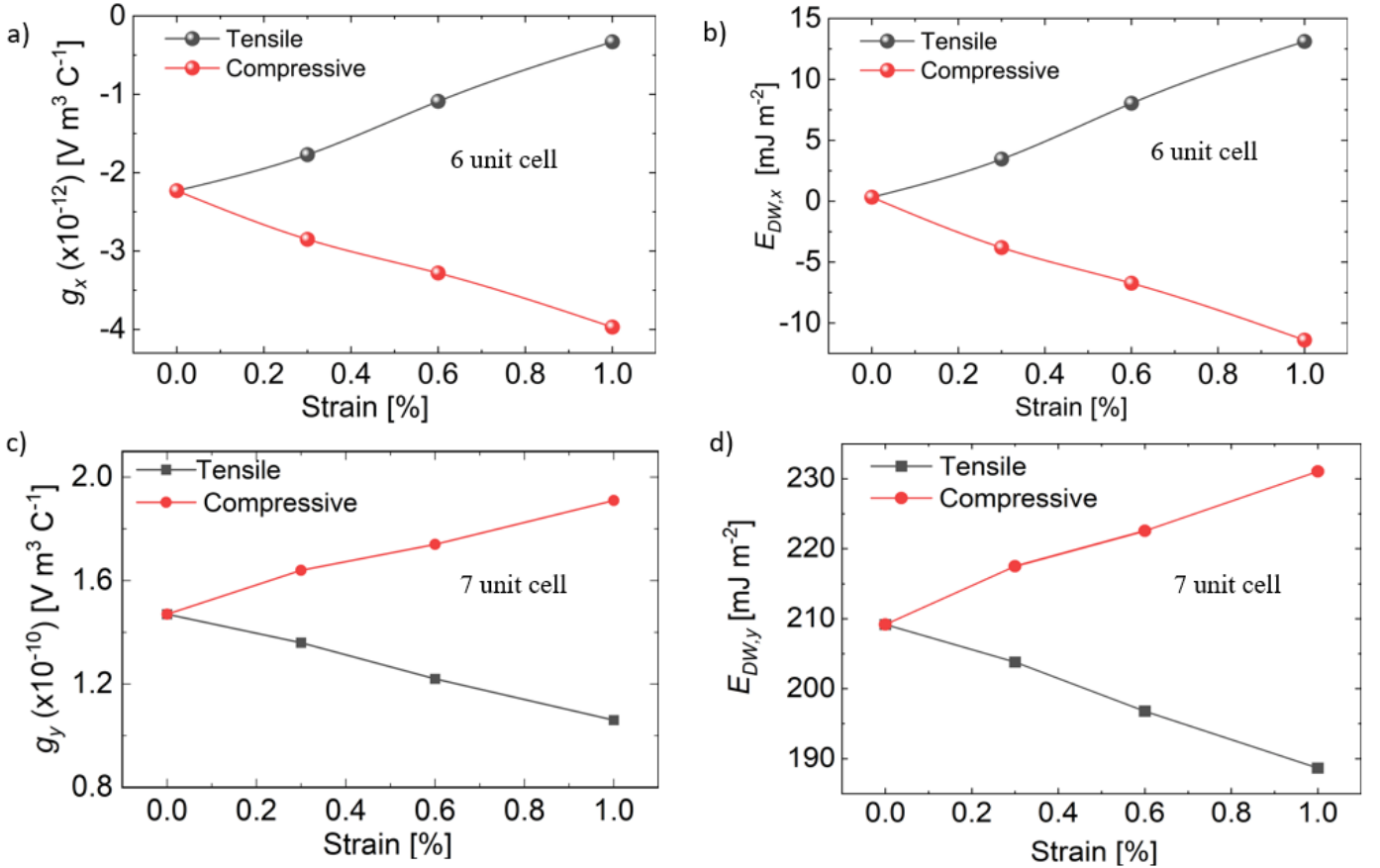

**Figure 10** a) Gradient energy variation with strain for APSL DW comprising 6-unit supercells. $g_x$ increases for tensile strain, but decreases for compressive strain. b) DW energy variation in APSL with tensile and compressive strain for 6-unit supercell. DW energy reduces (increases) with compressive (tensile) strain. This happens due to the induced compressive strain in APSL during the SD to MD transition. c) Gradient energy variation with tensile and compressive strain for CPL DW comprising 7-unit supercells. $g_y$ remains in the order of $10^{-10}$ V m$^3$ C$^{-1}$ up to 1% strain. d) DW energy variation in CPL with tensile and compressive strain for 7-unit supercell. The trends of $g_y$ and DW energy variation in CPL are opposite to the strain effect in APSL DW. This is due to the induced tensile strain in CPL during the SD to MD transition.

As discussed in Section 2.2.2, for the CPL configuration, the MD system feels slight tensile strain during the SD to MD transition. So, the application of tensile strain lowers the MD system energy, and then raises it after a certain tensile strain. But, SD system energy exhibits a monotonic increase in energy with an increase in tensile strain. Thus, tensile strain reduces DW energy. On the contrary, compressive strain deviates MD from its minimum energy configuration more than it deviates SD. Thus, compressive strain increases DW energy. This variation of DW energy is depicted in Figure 10d. Consequently, $g_y$ follows the same trend as the DW energy in this strain limit. Note that the trend of $g_x$ and $g_y$ variation is mutually opposite with respect to compressive and tensile strain. The captured effect of strain on $g_x$ and $g_y$ in addition to other parameters in the phase field model will be useful to analyze the impact of strain on the multidomain state of a sample.

## 3. Conclusion

In summary, in this work, we analyzed the orientation-dependence of lateral DWs in HZO and their gradient energy coefficients using first-principles calculations based on Density Functional Theory (DFT). Due to the unique alternate polar-spacer layer (APSL) configuration in one lateral direction, the corresponding polarization profile is atomically sharp near the domain wall. In the other lateral direction, continuous polar layer (CPL) configuration is observed, and polarization shows gradual variation near the DW. Along this direction, half-unit-cell polarization should be considered instead of unit-cell polarization. This is because domains can form with an odd number of half-unit cells and DW motion occurs in quanta of half-unit-cells. The energy barrier profile indicates that domain growth occurs faster along this direction compared to the former. We emphasize that due to this orientation-dependent DW characteristics, the gradient energy coefficient in HZO has a significant directional property. $g_x$ quantifies gradient energy between two consecutive oppositely polarized dipoles (unit cells) along the APSL direction. The magnitude of $g_x$ is small, negative, and in the order of $10^{-12}$ V m$^3$ C$^{-1}$. $g_y$ quantifies gradient energy between dipoles (half unit cells) along the CPL direction and is in the order of $10^{-10}$ V m$^3$ C$^{-1}$. $g_x$ and $g_y$ correspond to the most energetically stable systems of the two types of DWs and the order is valid over a wide range of domain widths. Compressive strain reduces $g_x$, but increases $g_y$. Similarly, tensile strain increases $g_x$, but reduces $g_y$. Both $g_x$ and $g_y$ affect the overall energetics of HZO samples. The results obtained in this work will give a comprehensive understanding of the gradient energy coefficient associated with lateral domain walls, which would be a crucial step for analyzing the multi-domain nature of HZO-based devices and their circuit applications.

## 4. Computational Section

For DFT calculation, Projector-Augmented Wave (PAW) pseudopotentials with non-linear core correction and Perdew–Burke–Ernzerhof generalized gradient approximation (GGA-PBE) as exchange-correlation functional are applied in Quantum Espresso software package. [27] All atomic structures shown in this work are generated using the Xcrysden software package. [28] To obtain the fully relaxed unit cell of HZO, $10^{-6}$ Rydberg (Ry) errors in ionic minimization energy and $10^{-4}$ Ry Bohr$^{-1}$ atomic force of all directional components are used, similar to previous works on HfO$_2$. [23] Kinetic energy cut-off of 60 Ry for wavefunction and 360 Ry for charge density and potential are used. $6 \times 6 \times 6$ Monkhorst-Pack grid of $k$-points is applied for Brillouin Zone sampling. MD periodic supercells of lateral (100) and (010) plane DWs are atomically relaxed at the SD lattice parameters using $1 \times 4 \times 4$ and $4 \times 1 \times 4$ Monkhorst-Pack grid of $k$-points, respectively. Gaussian smearing with spreading of 0.01 Ry is applied for better convergence. To accurately compare the energy of SD and MD supercells, SD supercells are calculated at the same $k$-point grid as MD.

FE polarization is calculated using the Berry phase approach of Kohn Sham states along with the modern theory of polarization to find both ionic and electronic contributions. [29] To find the energy barrier profiles of unit cell polarization reversal, as well as domain nucleation and DW motion, we use the Nudged Elastic Band (NEB) method. In strained configurations, to approximately mimic out-of-plane strain from the electrodes and in-plane stress from the substrate, we apply hydrostatic strain in our simulations. [23] Strains are calculated with respect to the volume of relaxed unstrained SD cells. All lattice parameters and atoms of SD cells are relaxed for each strain condition (both compressive and tensile). Then, the MD supercells are optimized at those lattice parameters, similar to what is done for configurations without strain.

*Acknowledgements*
This work was supported in part by Andrews Fellowship in Purdue University.

**References**

[1] V. Garcia, M. Bibes, Nat. Commun. 2014, 5, 4289.

[2] R. Yang, Nat. Electron. 2020, 3, 237.

[3] I. Chakrabortya, A. Jaiswal, A. K. Saha, S. K. Gupta, and K. Roy, Appl. Phys. Rev. 2020, 7, 021308.

[4] S. S. Cheema, D. Kwon, N. Shanker, R. Dos Reis, S.-L. Hsu, J. Xiao, H. Zhang, R. Wagner, A. Datar, M. R. Mccarter, C. R. Serrao, A. K. Yadav, G. Karbasian, C.-H. Hsu, A. J. Tan, L.-C. Wang, V. Thakare, X. Zhang, A. Mehta, E. Karapetrova, R. V. Chopdekar, P. Shafer, E. Arenholz, C. Hu, R. Proksch, R. Ramesh, J. Ciston, S. Salahuddin, Nature 2020, 580, 478.

[5] J. Müller, T. S. Böscke, U. Schröder, S. Mueller, D. Bräuhaus, U. Böttger, L. Frey, T. Mikolajick, Nano Lett. 2012, 12 (8), 4318.

[6] Y. Noh, M. Jung, J. Yoon, S. Hong, S. Park, B. S. Kang, S. E. Ahn, Curr. Appl. Phys. 2019, 19(4), 486.

[7] K. Jang, N. Ueyama; M. Kobayashi, T. Hiramoto, IEEE J. Electron Devices Soc. 2018, 6, 346.

[8] Y. L. Li, S. Y. Hu, Z. K. Liu, and L. Q. Chen, Appl. Phys. Lett. 2001, 78, 3878.

[9] V.L. Ginzburg, A.P. Levanyuk, J. Phys. Chem. Solids 1958, 6(1), 51.

[10] D. Zhao, T. Lenz, G. H. Gelinck, P. Groen, D. Damjanovic, D. M. De Leeuw, I. Katsouras, Nat. Commun 2019, 10, 2547.

[11] S. Matzen, S. Fusil, C. R. Phys. 2015, 16(2), 227.

[12] A. K. Saha, S. K. Gupta, J. Appl. Phys. 2021, 129, 080901.

[13] A. K. Saha, M. Si, K. Ni, S. Datta, P. D. Ye and S. K. Gupta, In 2020 IEEE International Electron Devices Meeting (IEDM), San Francisco, CA 2020, 4.3.1.

[14] P. Zubko, J. C. Wojdeł, M. Hadjimichael, S. F.-Pena, A. Sené, I. Luk'yanchuk, J.-M. Triscone, J. Íñiguez, Nature 2016, 534, 524.

[15] M. Hoffmann, M. Gui, S. Slesazeck, R. Fontanini, M. Segatto, D. Esseni, T. Mikolajick, Adv. Func. Mat. 2022, 32(2), 2108494.

[16] Y. -H. Shin, I. Grinberg, I.-W. Chen, A. M. Rappe, Nature 2007, 449, 881.

[17] O. Diéguez, M. Stengel, Phys. Rev. X 2022, 12, 031002.

[18] H.-J. Lee, M. Lee, K. Lee, J. Jo, H. Yang, Y. Kim, Sci. 2020, 369, 1343.

[19] T. Kiguchi, T. Shiraishi, T. Shimizu, H. Funakubo, T. J. Konno, Jpn. J. Appl. Phys. 2018, 57, 11UF16.

[20] W. Ding, Y. Zhang, L. Tao, Q. Yang, Y. Zhou, Acta Mater. 2020, 196, 556.

[21] T. D. Huan, V. Sharma, G. A. Rossetti, Jr., and R. Ramprasad, Phys. Rev. B 2014, 90, 064111.

[22] R. Materlik, C. Künneth, and A. Kersch, J. Appl. Phys., 2015, 117, 134109.

[23] M. Dogan, N. Gong, T. P. Maae, S. I. Beigi, Phys. Chem. Chem. Phys. 2019, 21, 12150.

[24] D. -H. Choe, S. Kim, T. Moon, S. Jo, H. Bae, Mater. Today 2021, 50, 8.

[25] A. N. Morozovska, E. A. Eliseev, J. J. Wang, G. S. Svechnikov, Y. M. Vysochanskii, V. Gopalan, and L.-Q. Chen, Phys. Rev. B 2010, 81, 195437.

[26] T. S. Böscke, J. Müller, D. Bräuhaus, U. Schröder, U. Böttger, Appl. Phys. Lett. 2011, 99, 102903.

[27] P. Giannozzi, S. Baroni, N. Bonini, M. Calandra, R. Car, C. Cavazzoni, D. Ceresoli, G. L. Chiarotti, M. Cococcioni, I. Dabo, A. Dal Corso, S. de Gironcoli, S. Fabris, G. Fratesi, R. Gebauer, U. Gerstmann, C. Gougoussis, A. Kokalj, M. Lazzeri, L. Martin-Samos, N. Marzari, F. Mauri, R. Mazzarello, S. Paolini, A. Pasquarello, L. Paulatto, C. Sbraccia, S. Scandolo, G. Sclauzero, A. P. Seitsonen, A. Smogunov, P. Umari, and R. M. Wentzcovitch, J. Phys.: Condens. Matter 2009, 21, 395502.

[28] A. Kokalj, Comp. Mater. Sci. 2003, 28, 155-168.

[29] N. A. Spaldin, J. of Solid State Chem. 2012, 195, 2.

Tanmoy Kumar Paul*, Atanu Kumar Saha, Sumeet Kumar Gupta
*Purdue University, West Lafayette, IN, 47907, USA*
*Email: paul115@purdue.edu*

| U | | | | $D_a$ | | | | $D_b$ | | | |
|---|---|---|---|---|---|---|---|---|---|---|---|
| Zr | 0.016 | 0.032 | 0.000 | Zr | 0.016 | -0.032 | 0.000 | Zr | 0.016 | 0.032 | 0.000 |
| Zr | 0.484 | 0.532 | 0.000 | Zr | 0.484 | 0.468 | 0.000 | Zr | 0.484 | 0.532 | 0.000 |
| O | 0.821 | 0.372 | 0.147 | O | 0.689 | 0.357 | 0.358 | O | 0.811 | 0.143 | 0.358 |
| O | 0.679 | 0.872 | 0.147 | O | 0.811 | 0.857 | 0.358 | O | 0.689 | 0.643 | 0.358 |
| O | 0.287 | 0.228 | 0.253 | O | 0.216 | 0.227 | 0.253 | O | 0.284 | 0.273 | 0.253 |
| O | 0.213 | 0.728 | 0.253 | O | 0.284 | 0.727 | 0.253 | O | 0.216 | 0.773 | 0.253 |
| Hf | 0.484 | -0.032 | 0.500 | Hf | 0.484 | 0.032 | 0.500 | Hf | 0.484 | -0.032 | 0.500 |
| Hf | 0.016 | 0.468 | 0.500 | Hf | 0.016 | 0.532 | 0.500 | Hf | 0.016 | 0.468 | 0.500 |
| O | 0.689 | 0.643 | 0.642 | O | 0.821 | 0.627 | 0.853 | O | 0.679 | 0.873 | 0.853 |
| O | 0.811 | 0.143 | 0.642 | O | 0.679 | 0.127 | 0.853 | O | 0.821 | 0.372 | 0.853 |
| O | 0.284 | 0.273 | 0.747 | O | 0.213 | 0.271 | 0.747 | O | 0.287 | 0.229 | 0.747 |
| O | 0.216 | 0.773 | 0.747 | O | 0.287 | 0.771 | 0.747 | O | 0.213 | 0.729 | 0.747 |

**Figure S1** Fractional atomic coordinates of U, $D_a$ and $D_b$ mentioned in Figure 3. $D_c$ and $D_d$ are derived by altering the sequence of polar and spacer layer in $D_a$ and $D_b$ respectively. This is obtained by shifting $D_a$ and $D_b$ respectively by half unit cell and folding the atoms within the unit cell.

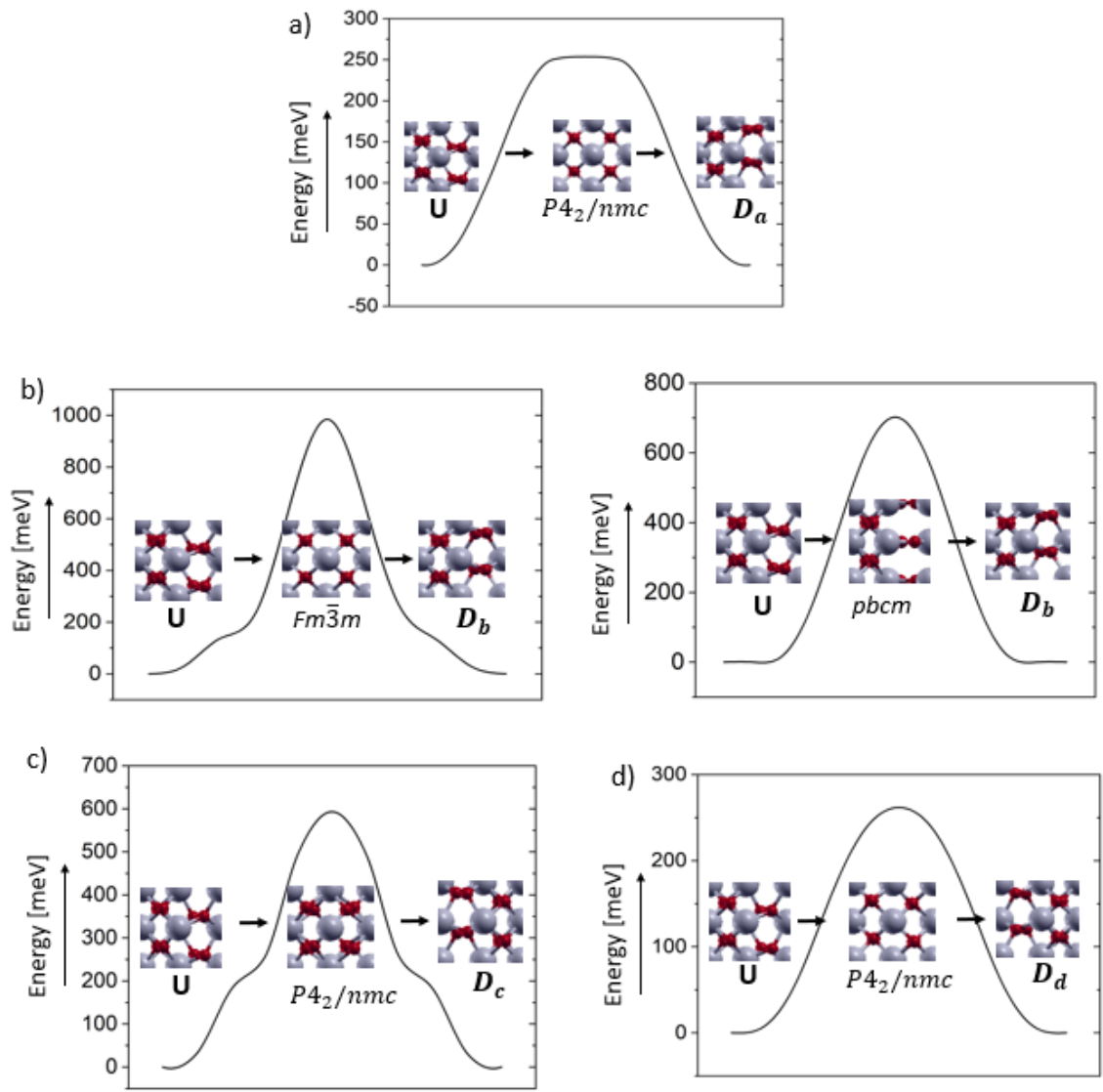

**Figure S2** a)-d) Polarization reversal path for upward polarized unit cell, U to downward polarized unit cells $D_a$, $D_b$, $D_c$ and $D_d$ respectively. The unit cells are taken from Figure 3 of main text. From U to $D_b$ transition, both $Fm\bar{3}m$ and $pbcm$ can occur as intermediate state as shown in the two paths of b). While passing through $Fm\bar{3}m$ phase, the non-centrosymmetric O atoms shift without crossing the Hf/Zr plane. On the contrary, while passing through $pbcm$ phase, the non-centrosymmetric O atoms cross the Hf/Zr plane. For U to $D_a$, $D_c$ and $D_d$ transition, $P4_2/nmc$ occurs as the intermediate phase.

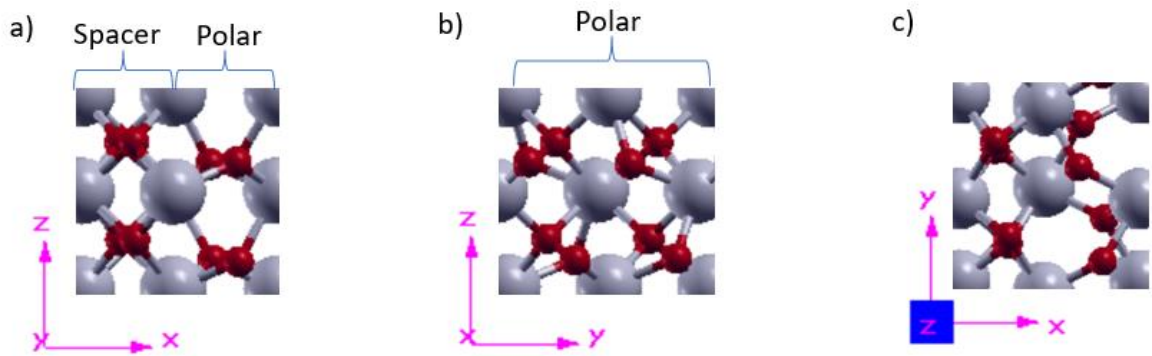

**Figure S3** a)-c) Atomic Configuration of HZO unit cell observed from three orthogonal planes. Along x direction, there are alternate polar and spacer layers. Along y direction, all layers are polarized.

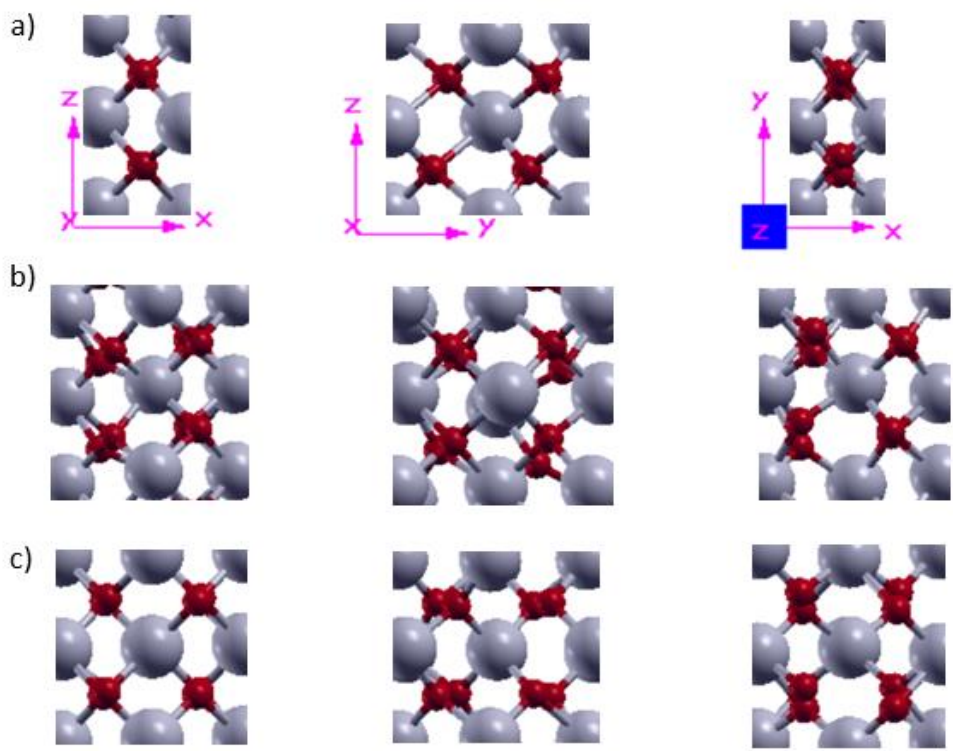

**Figure S4** a)-c) Atomic configurations of APSL DWs of supercells, $MD_{x,a}$, $MD_{x,c}$ and $MD_{x,d}$ respectively. The DWs resemble tetragonal, distorted orthorhombic and tetragonal phases respectively. DW of a) is half unit cell wide and so it shows half-cell wide configuration. DW of supercell, $MD_{x,b}$ is identical to the original orthorhombic phase and is not shown here.

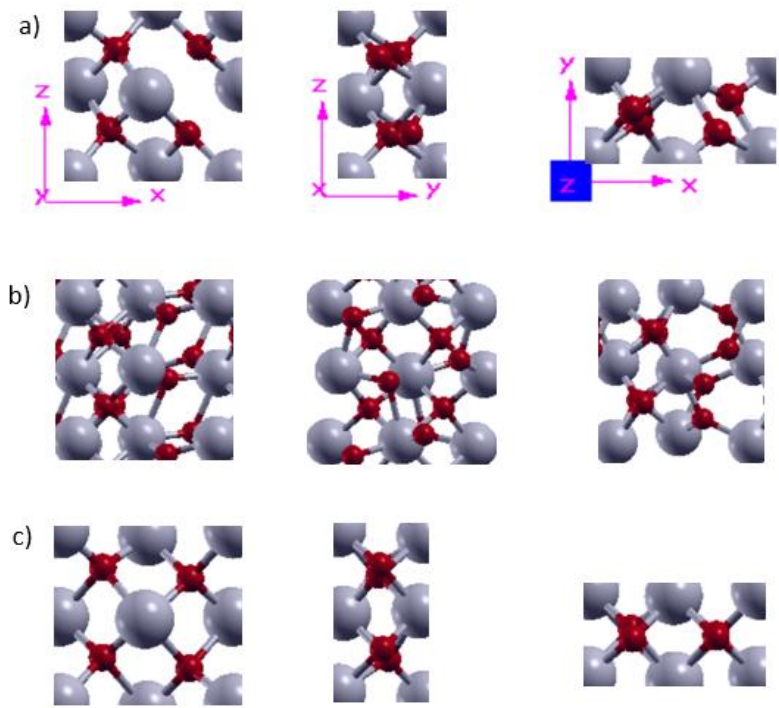

**Figure S5** a)-c) Atomic configurations of CPL DWs of supercells, $MD_{y,a}$, $MD_{y,b}$ and $MD_{y,c}$ respectively. The atomic configurations near DW resemble distorted tetragonal, *pbcm* and distorted tetragonal respectively. a) and c) show half-cell wide configuration at the center of the DWs.

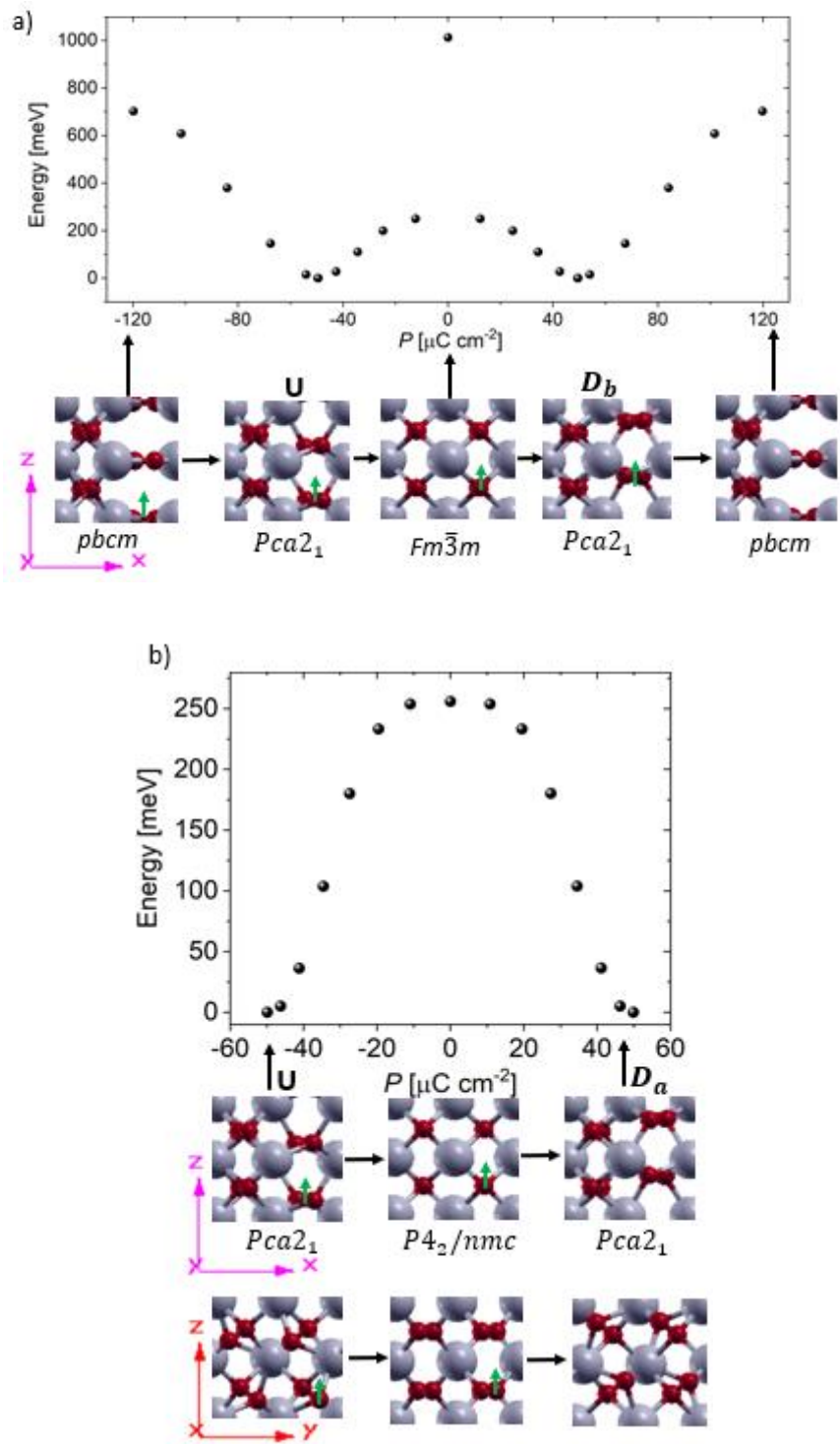

**Figure S6** a) Energy barrier profile for half unit cell shifting of oxygen atoms. The $Pca2_1$ phases here are unit cell U and $D_b$. The curve is obtained from merging the two polarization reversal paths of Figure S1b. Local free energy of dipoles near APSL DW is extracted from this curve. b) Energy barrier profile from unit cell U to unit cell $D_a$. As the two neighboring half-unit cells along [010] direction are symmetric, local free energy of dipoles (half-unit cell) in CPL configuration is extracted by dividing the barrier energies by 2. Polarization is calculated from modern theory of polarization where $Fm\bar{3}m$ and $P4_2/nmc$ phases are the non-polar phases with zero polarization in a) and b) respectively. The green arrow shows the direction of the shift of O atoms.

a)

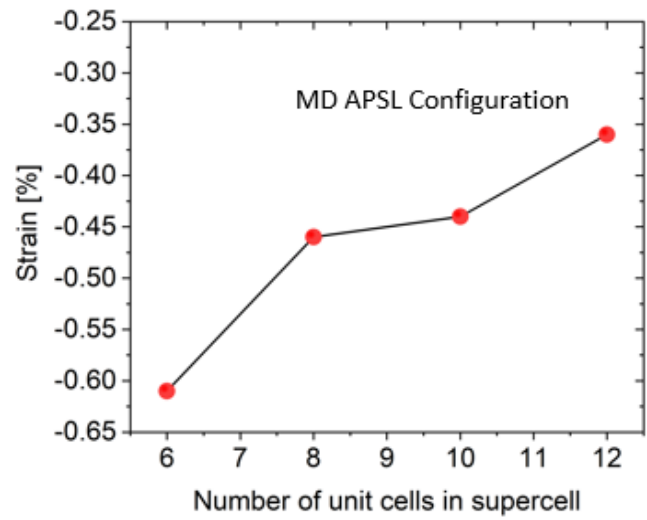

b)

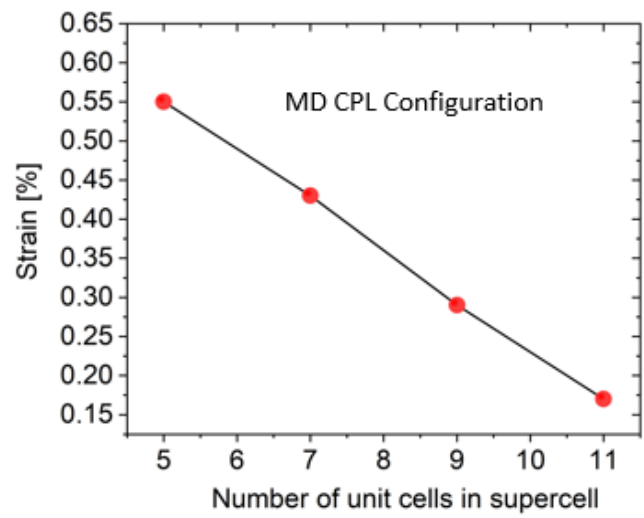

**Figure S7** a)-b) Hydrostatic strain induced during transition from SD to MD for APSL and CPL respectively. Hydrostatic strain is calculated from unconstrained MD volume and unconstrained SD volume as the following:

$$Strain = \frac{Vol_{MD} - Vol_{SD}}{Vol_{SD}} \times 100\%$$

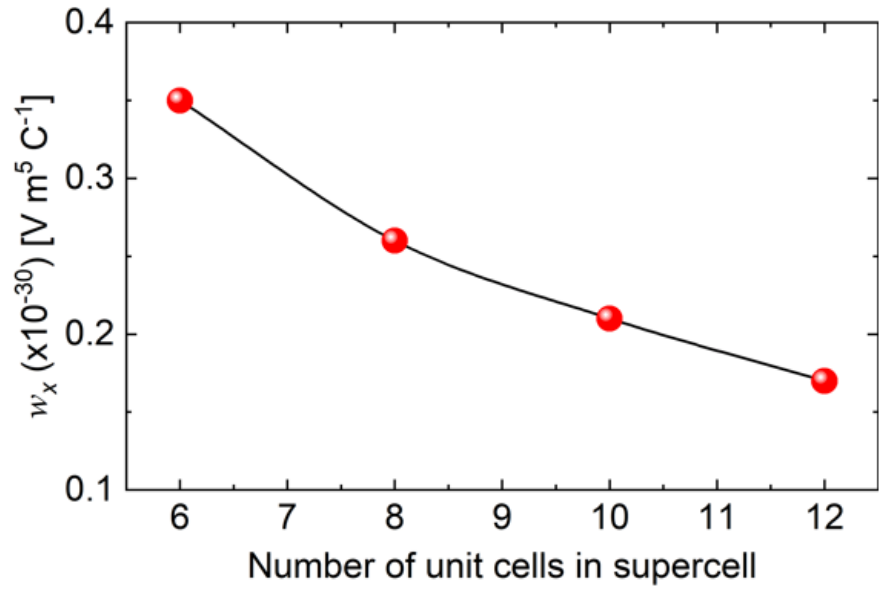

**Figure S8** Higher order gradient energy coefficient, $w_x$ along APSL direction for different domain width. This is obtained by equating $f_{grad,high}$ to the excess local free energy at the two lattice points near the sharp DW as follows.

$$E_{ex} = E - E_{min} = \sum w_x \left(\frac{P_{i-1} - 2P_i + P_{i+1}}{\Delta x^2}\right)^2$$

Here, $E_{ex}$= excess local free energy at only those two points near the DW, $E_{min}$= minima of double well potential barrier, E= local free energy at the two points

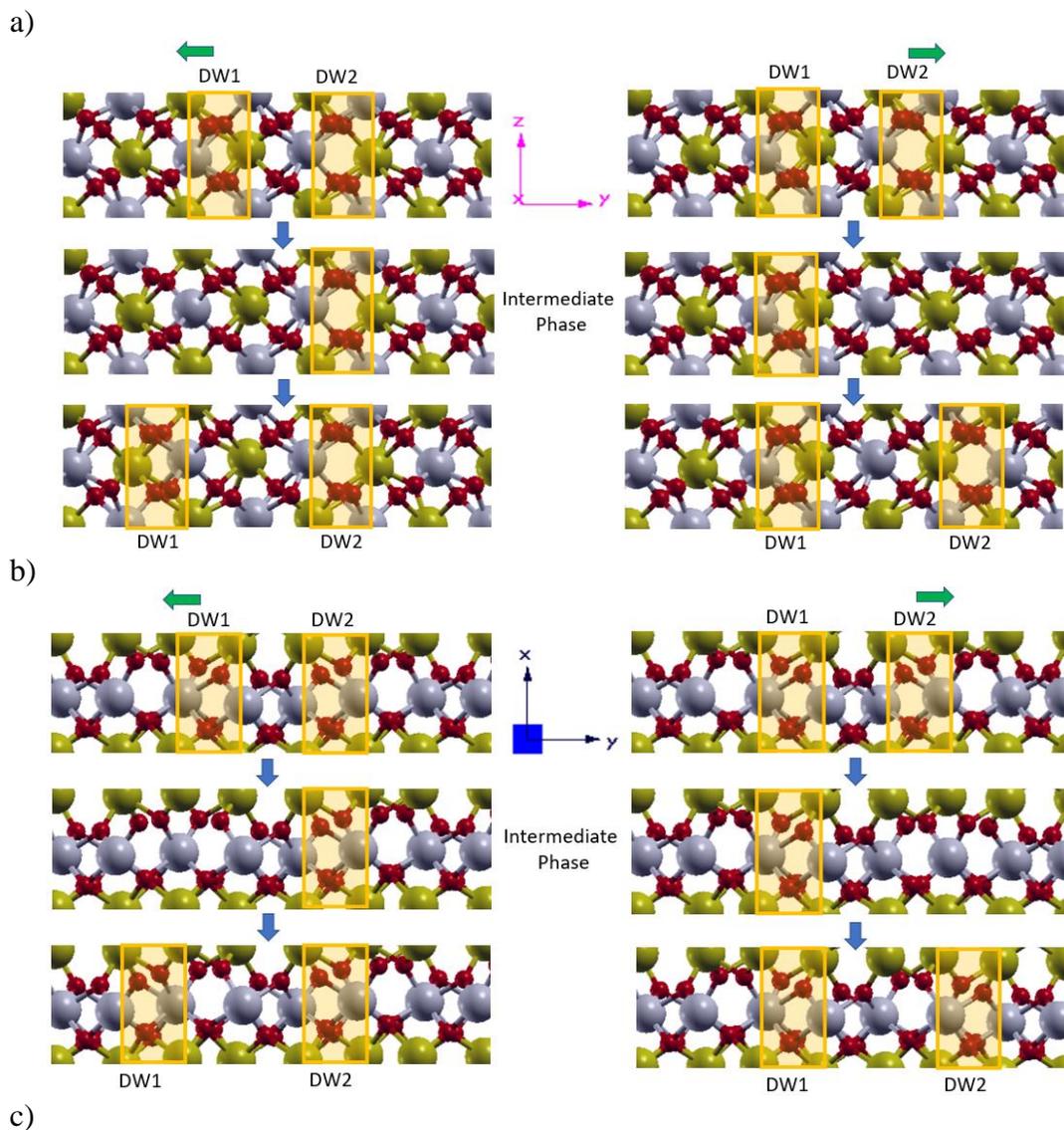

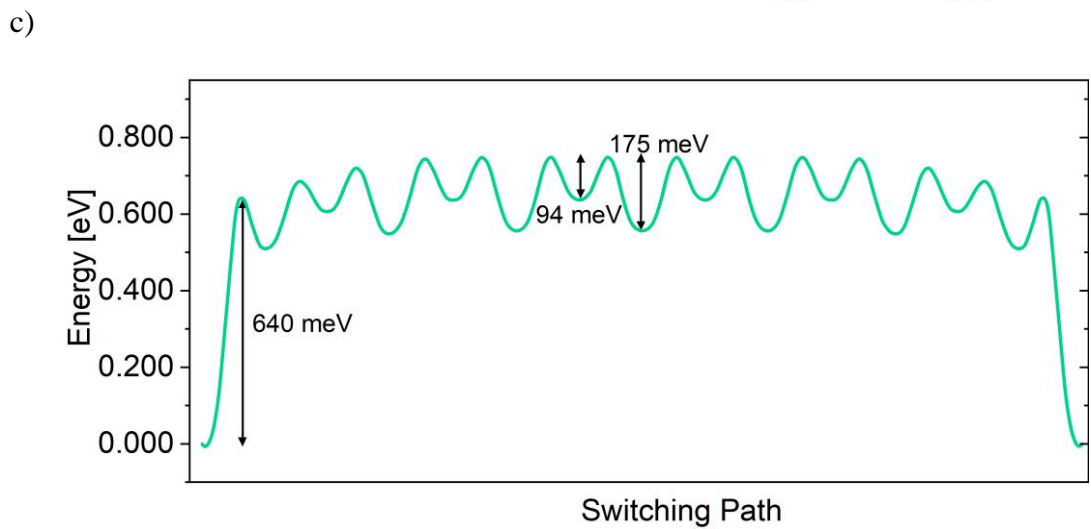

**Figure S9** DW shift just after nucleation and corresponding intermediate phases for CPL with (100) monoatomic plane. Left panels of a-b: The domain growth mechanism when DW1 shifts towards left keeping DW2 fixed. Right panels of a-b: The domain growth mechanism when DW2 shifts towards right keeping DW1 fixed. The middle image of each path is the intermediate image with the highest energy. c) Energy Barrier profile for domain nucleation and growth. The planes of DW1 and DW2 that shift have mixture of Hf and Zr atoms and relative arrangement of Hf and Zr atoms in the two DWs are same. This causes identical heights of energy barrier for the two DW motions. Here, Hf (silver atoms) and Zr (yellow atoms) are distinguished by different colors to illustrate the difference properly.

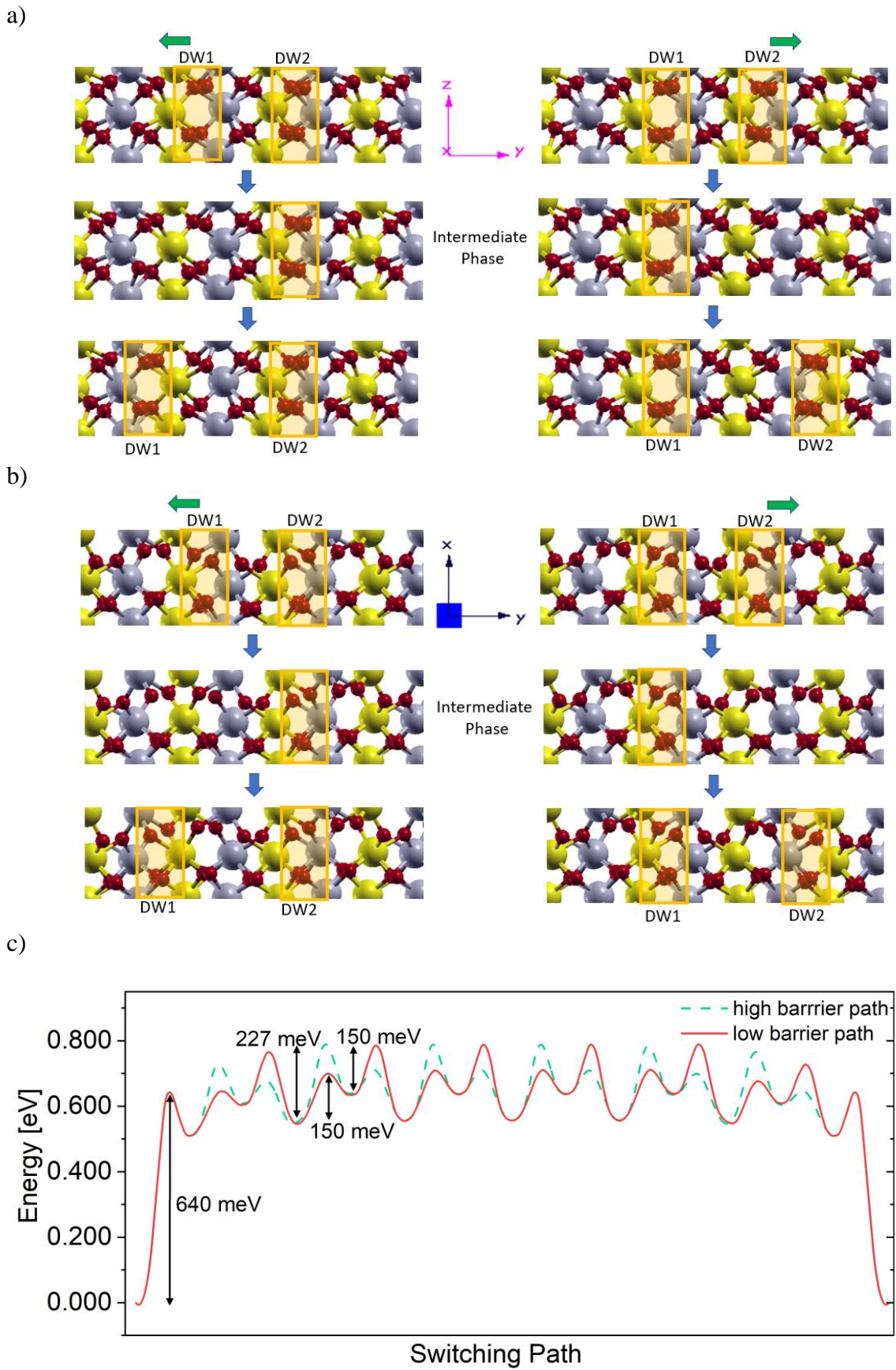

**Figure S10** DW shift just after nucleation and corresponding intermediate phases for CPL with (010) monoatomic plane. Left panels of a-b: The domain growth mechanism when DW1 shifts towards left keeping DW2 fixed. Right panels of a-b: The domain growth mechanism when DW2 shifts towards right keeping DW1 fixed. The middle image of each path is the intermediate image with the highest energy. c) Energy Barrier profile for domain nucleation and growth. Different monoatomic planes (of Hf and Zr) shift for DW1 and DW2, which causes significantly different heights of energy barrier for the two DW motions. Here, Hf (silver atoms) and Zr (yellow atoms) are distinguished by different colors to illustrate the difference properly.

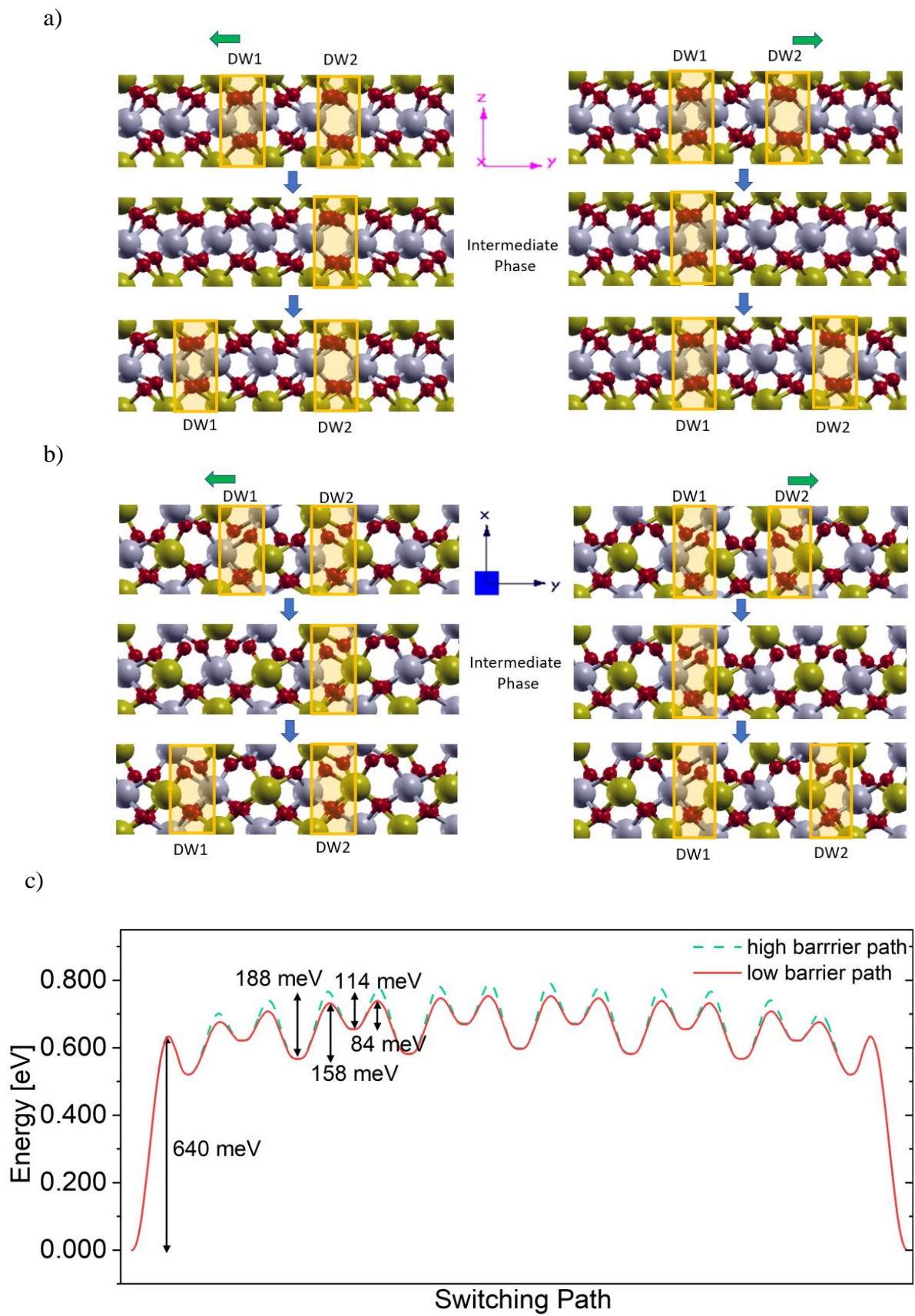

**Figure S11** DW shift just after nucleation and corresponding intermediate phases for CPL with (001) monoatomic plane. Left panels of a-b: The domain growth mechanism when DW1 shifts towards left keeping DW2 fixed. Right panels of a-b: The domain growth mechanism when DW2 shifts towards right keeping DW1 fixed. The middle image of each path is the intermediate image with the highest energy. c) Energy Barrier profile for domain nucleation and growth. Although the planes of DW1 and DW2 that shift have mixture of Hf and Zr atoms, relative arrangement of Hf and Zr atoms in the two DWs are different. This causes slightly different heights of energy barrier for the two DW motions. Here, Hf (silver atoms) and Zr (yellow atoms) are distinguished by different colors to illustrate the difference properly.